# A relation between the dark mass of elliptical galaxies and their shape

A. Deur★†
*University of Virginia, Charlottesville, VA 22903, USA*

## ABSTRACT

We have studied a large number of elliptical galaxies and found a correlation between their dark matter content and the ellipticity of their visible shape. The galaxies were strictly selected so that only typical medium-size elliptical galaxies were considered. Galaxies with unusual characteristics were rejected to minimize point-to-point data scatter and avoid systematic biases. Data from six different techniques of extracting the galactic dark matter content were used to avoid methodological biases. A thorough investigation of the interrelation between attributes of elliptical galaxies was carried out to assess whether the correlation originates from an observational bias, but no such origin could be identified. At face value, the correlation found implies that at equal luminosities, rounder medium-size elliptical galaxies appear to contain less dark matter than flatter elliptical galaxies, e.g. the rounder galaxies are on average four times less massive than the flatter ones. This is puzzling in the context of the conventional model of cosmological structure formation.

## 1 INTRODUCTION

Dark matter cosmology has been successful at determining many large-scale features of the Universe (Komatsu et al. 2011) but open questions remain at the galactic and semigalactic scales (Napolitano et al. 2009; Auger et al. 2010). A search for relationships between the observed luminous matter of a galaxy and its dark matter content can potentially advance our understanding of dark matter cosmology at this scale. Useful and intriguing empirical correlations have been found, for example the Tully–Fisher relation (Tully & Fisher 1977) for spiral galaxies, which relates their rotation speed to their absolute luminosity, or for elliptical galaxies, the Fundamental Plane (Djorgovski 1987; Dressler et al. 1987), which includes the Faber–Jackson (Faber & Jackson 1976) and Kormendy (Kormendy 1977) relations and links their effective radius $R_{\mathrm{eff}}$, their surface brightness and the statistical dispersion of stellar velocities $\sigma$. However, none of these correlations involve directly the dark matter content of galaxies. There is also the puzzling observation of little dark matter content in some elliptical galaxies (Romanowsky et al. 2003). We report here on a search for a correlation between the most obvious visible characteristic of elliptical galaxies, the ellipticity, on which their Hubble classification is solely based, and their relative amount of dark matter, expressed as the galactic total mass (dark+luminous) normalized to luminosity ($M/L$).

★ E-mail: deurpam@jlab.org
† Present Address: Thomas Jefferson National Accelerator Facility, Newport News, VA 23606, USA.

## 2 METHOD AND CHOICE OF DATA

Elliptical galaxies generally have triaxial ellipsoidal shapes with ellipticities depending on radii. However, to first approximation the galactic shapes can be simply parametrized as oblate ellipsoids of constant ellipticities, $\varepsilon$. (This approximation has been used by many of the publications used in this work. The ellipticity $\varepsilon$ is defined as $1 - R$, with $R$ being the minor to major axis ratio of the ellipsoid.) These span a large range, from spheres ($\varepsilon = 0$) to flattened ellipsoids ($\varepsilon = 0.7$), providing us with a group of smoothly varying shapes. However, only the projection of $\varepsilon$ on our observation plane (the apparent ellipticity, $\varepsilon_{\mathrm{app}}$) can be measured, the true ellipticity ($\varepsilon_{\mathrm{true}}$) remaining elusive. In addition, the $M/L$ ratio for elliptical galaxies is hard to measure accurately (Romanowsky et al. 2003). To overcome these difficulties, we select large and homogeneous samples of elliptical galaxies for which $M/L$ has been extracted using different methods. Effects of galaxy peculiarities are minimized by a homogeneity requirement and suppressed statistically. The projection problem is addressed statistically. We select 685 galaxies from 42 publications. This provides 41 homogeneous samples. (We combined publications from the same group of authors that used the same method, and sometime one publication provides several determinations of $M/L$. So our number of samples is different from the number of publications used. Also, many of the 685 data points correspond to same galaxies. We have a total of 255 different galaxies.) Each publication provides $M/L$ (or related ratios) for at least several galaxies, allowing us to study subgroups of data with consistent $M/L$ extraction. We also studied 24 other such publications but did not use their results for one of the following reasons.

(i) The data have been superseded by newer data from the same author group on the same sample of galaxies (Grillo et al. 2008; Barnabè et al. 2009).

(ii) Once our selection criteria have been applied (see the next section), the publication has not enough remaining galaxies for a meaningful study (less than three suitable galaxies; Saglia, Bertin & Stiavelli 1992; Saglia et al. 1993; Bertin et al. 1994; Carollo & Danziger 1994; Brighenti & Mathews 1997; Trinchieri, Fabbiano & Kim 1997; Fassnacht & Cohen 1998; Gebhardt et al. 2003; Holden et al. 2005; Humphrey et al. 2005, 2006; Napolitano et al. 2005; O'sulivan, Sanderson & Ponman 2007; Cappellari et al. 2009; Das et al. 2010; More et al. 2011).

(iii) The publication did not explicitly provide the values for the dark matter content (Carollo et al. 1995; van Dokkum & Stanford 2003; di Serego Alighieri et al. 2005; Coccato et al. 2009; Proctor et al. 2009; Conroy & van Dokkum 2012).

## 3 GENERAL SELECTION CRITERIA

To study a correlation revealing intrinsic general properties of elliptical galaxies, one needs to select typical elliptical galaxies. This avoids diluting (by including effects particular to a galaxy) or biasing any possible correlation (via systematic effects from a class of galaxies, e.g. compact or giant elliptical galaxies). With this in mind, from each publication we select a sample of elliptical galaxies as homogeneous as possible. Only medium-size elliptical galaxies are considered. These tend to have nearly isotropic random velocities (contrary to e.g. giant elliptical galaxies). We require undisturbed galaxies to avoid galaxy–galaxy interaction from invalidating the method used to extract $M/L$ (e.g. virial theorem, equations of hydrostatic equilibrium or strong lensing equations). The strict selection applied to samples of local galaxies must be relaxed for samples of distant galaxies, since those are not as well characterized. (Distant galaxies are typically considered in strong lensing studies or in studies of the Fundamental Plane time evolution.) The local and distant rejection criteria are as follows [identification and numbers are from either NASA/IPAC Extragalactic Database (NED) or the publication from which $M/L$ originates].

### 3.1 Classes of rejected local galaxies

(i) Lenticular galaxies (S0-type), transition type (E+) and spiral galaxies, since they belong to different classes of galaxies.

(ii) Galaxies having an active galactic nucleus (AGN), since, for mature galaxies, it may reveal that the galaxies have been disturbed recently.

(iii) Seyfert galaxies and BL Lacertae objects, for the same reasons as AGN.

(iv) Peculiar galaxies and galaxies listed in the Arp Atlas of Peculiar Galaxies (Arp 1966), since they show clear signs of disturbance.

(v) Galaxies with H II emission, since the presence of H II regions is peculiar for elliptical galaxies. Furthermore, it may reveal a recent disturbance. It may also bias $M/L$, since newly formed blue stars increase the luminosity.

(vi) Galaxies with low-ionization nuclear emission-line regions (LINER), since they may be due to AGN or star births, in both cases a sign that the elliptical galaxies have been disturbed.

(vii) Compact elliptical galaxies since they belong to a different class of elliptical galaxies.

(viii) Supergiant elliptical galaxies (cD), giant elliptical galaxies (D), brightest cluster galaxies (BrClG), since they belong to different classes of elliptical galaxies. Furthermore, giant elliptical galaxies are characterized by more anisotropic random velocities and tend to be triaxial. In addition, the environment of the cluster or group to which these galaxies belong may contribute to the determination of the galaxy dark matter content (Fukazawa et al. 2006; Nagino & Matsushita 2009).

(ix) E? galaxies. We assume that this lack of nomenclature knowledge reflects poor measurements and may contaminate the sample with non-elliptical galaxies.

(x) Bright X-ray and very faint X-ray galaxies. Very faint X-ray galaxies show signs of disturbed hydrostatic equilibrium (Fukazawa et al. 2006; Nagino & Matsushita 2009). In the case of bright X-ray galaxies, the hot gas property merges with the extragalactic gas of the group/cluster, indicating some influence from the environment and a contribution of the group/cluster dark matter to the galactic $M/L$ (Fukazawa et al. 2006; Nagino & Matsushita 2009). In addition, these galaxies tend to be cD, D or BrClG.

We keep low-excitation radio galaxies and weak emission-line radio galaxies as we have seen no reason to reject them.

### 3.2 Classes of rejected distant galaxies

(i) Massive galaxies, typically with $M \geq 5 \times 10^{11}$ $M_\odot$, to reduce the amount of cD, D or BrClG galaxies.

(ii) Galaxies with velocity dispersions $\sigma \leq 225$ km s$^{-1}$. This criterion is applied only if no distinction is made in the publication between S0 and elliptical galaxies, or if the classification is not reliable enough. This is to suppress possible S0 contamination, since S0 tend to have $\sigma \leq 225$ km s$^{-1}$.

The two lists above describe our standard selection criteria. When the data are not recent, we require in addition that the galactic characteristics given in the publication agree with their values from NED. In some cases, when statistics, i.e. the number of galaxies, is low, we relaxed the selection criteria (typically keeping LINER galaxies). Unless such relaxation is justified (e.g. the method does not require the galaxy to be in equilibrium), the analysis is considered to be of lower reliability. This, together with the smaller statistics, makes these analyses contribute little to the final result. When such variations from our standard criteria are applied, we documented them in Table A1 in the appendix.

We remark that the rejection criteria have been chosen before performing the analysis. In this sense, the analysis has been blind and is as free of subjective bias as possible.

## 4 METHODS FOR THE DARK MATTER CONTENT EXTRACTION

We use data from six different categories of methods employed to deduce $M/L$. This mitigates the possibility of a systematic methodological bias. These are as follows.

(i) Virial theorem. We employ virial data from eight publications (Bacon, Monnet & Simien 1985; Lauer 1985; Bender et al. 1989; Prugniel & Simien 1996; Kelson et al. 2000; van der Wel et al. 2005; Rettura et al. 2006; Leier 2009). For cases where published analyses do not allow for ellipticity, we correct the $M/L$ using the tensor expression given in Bacon et al. (1985). [This important correction derived analytically was experimentally verified in van der Marel (1991).]

(ii) Stellar dynamics modelling. We use 11 publications (van der Marel 1991; Magorrian et al. 1998; Kronawitter 2000; Cappellari et al. 2006; Thomas et al. 2007; van der Marel & van Dokkum 2007a, b; Thomas et al. 2011; Wegner et al. 2012; Cappellari et al. 2013a,b). An advantage of this method is that $\varepsilon_{\rm true}$ can be inferred and used directly in calculations.

(iii) Interstellar gas X-ray emission data. The two publications we used (Fukazawa et al. 2006; Nagino & Matsushita 2009) assume spherical symmetry. We partly accounted for ellipticity by replacing Newton's shell theorem (used in the derivation of the equation of hydrostatic equilibrium) with a relation including ellipticity. This yields a correction numerically similar to the tensor correction (Bacon et al. 1985) to the scalar virial theorem.

(iv) Planetary nebulae and globular cluster data. The four publications used here (Capaccioli, Napolitano & Arnaboldi 1992; Magorrian & Ballantyne 2001; Romanowsky et al. 2003; Deason et al. 2012) assume spherical symmetry. To account for ellipticity, we note that the ellipticity corrections to the virial and shell theorems are close and we assume the same correction.

(v) Gas disc dynamics. We use three publications (Bertola et al. 1991, 1993; Pizzella et al. 1997).

(vi) Strong lensing. We use 16 publications (Jackson et al. 1998; Keeton, Kochanek & Falco 1998; Treu & Koopmans 2004; Ferreras, Saha & Williams 2005; Koopmans et al. 2006; Jiang & Kochanek 2007; Ferreras, Saha & Burles 2008; Cardone et al. 2009, 2011; Grillo et al. 2009; Leier 2009; Auger et al. 2010; Barnabè et al. 2011; Faure et al. 2011; Leier et al. 2011; Ruff et al. 2011). Effects of $\varepsilon$ are expected to be small (Grillo et al. 2009) and no correction is applied.

Clearly, from such varied methods of determining $M/L$, different correlations with ellipticity can result depending on the treatment of $\varepsilon$, systematic modelling biases or sampling biases. For example, strong lensing tends to sample more massive galaxies, whose corresponding $M/L(\varepsilon)$ correlation strength might differ from those of medium-mass galaxies. It is to minimize such biases that we sought to employ all available methods.

## 5 CORRELATION ANALYSIS

Linear fits of $M/L$ to $\varepsilon$ are carried out for each of the 41 homogeneous samples. The fit results are then carefully combined (see Section 7), which reveals a significant correlation between $M/L$ and $\varepsilon$. Results from 4 out of the 41 samples are shown on Fig. 1. The best fits for all 41 samples are given in Appendix A and the corresponding figures in Appendix B. As will be shown in Section 6, the large scatter of the data points within a given sample, see e.g. the top-left panel of Fig. 1, results from the random projection of the elliptical galaxies on our observation plane. All slopes in Fig. 1 indicate a positive correlation, with 4.4, 4.3, 1.3 and 1.2$\sigma$ deviation from zero, from top-left to bottom-right panels, respectively. (Fits of $M/L$ without accounting for $\varepsilon$, i.e. using straight horizontal lines, yield $\chi^2/ndf$ values that are 1.5, 1.3, 1.1 and 1.1 time larger, respectively.)

## 6 PROJECTION CORRECTION

A difficulty with elliptical galaxies is that we observe only their projection, with their true ellipticities being unknown. However, the orientation of galaxies with respect to us is random. Hence, under minimal assumptions, the $\varepsilon_{\rm true}$ distribution can be computed from the $\varepsilon_{\rm app}$ distribution and the projection corrected on a statistical basis. This correction is independent of the $M/L$ extraction method if the ellipticity distributions in the galaxy samples are the same. We assume so and apply an identical correction to all the data sets (except for those already using $\varepsilon_{\rm true}$, as for example in the bottom-right plot of Fig. 1). We use the $\varepsilon_{\rm app}$ distribution and $M/L$ versus $\varepsilon_{\rm app}$ correlation obtained from Bacon et al. (1985) to estimate the correction because it is the second largest sample of galaxies [the largest sample is from Prugniel & Simien (1996) but the authors indicate it may be slightly biased]. The relation between $\varepsilon_{\rm true}$ and $\varepsilon_{\rm app}$ for an oblate spheroid viewed at an angle $i$ is

$$\varepsilon_{\rm app} = 1 - \sqrt{(1-\varepsilon_{\rm true})^2 \sin^2 i + \cos^2 i}. \quad (1)$$

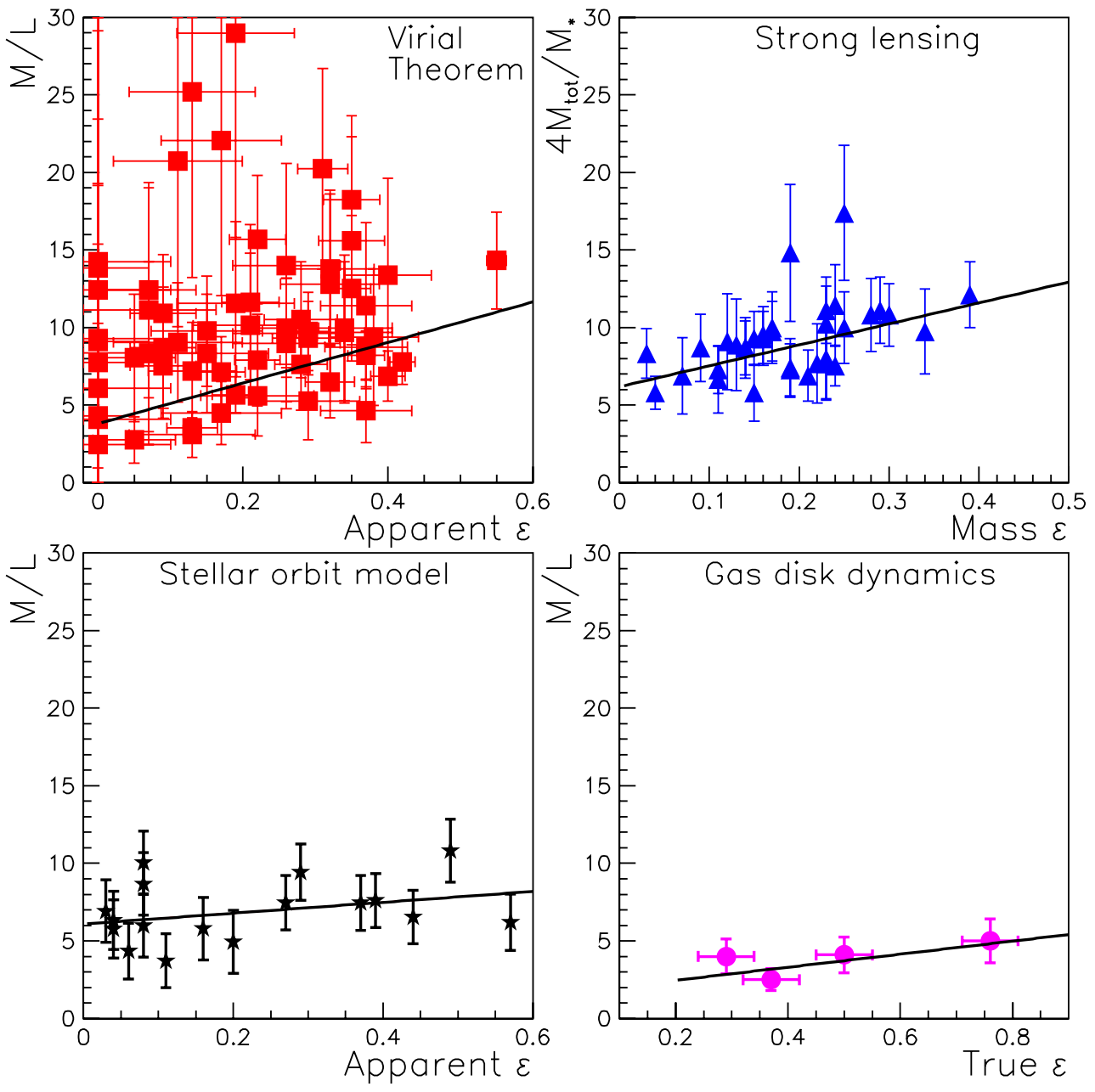

**Figure 1.** Example of galactic mass over luminosity in solar $M_\odot/L_\odot$ units versus ellipticity from four publications using different methods. Top left: virial theorem (Bacon et al. 1985). Top right: strong lensing (Auger et al. 2010) (we plot four times the total mass over the stellar mass $4M_{tot}/M_* \simeq M/L$). Bottom left: stellar orbit modelling (van der Marel & van Dokkum 2007b). Bottom right: gas disc dynamics (Bertola et al. 1991, 1993).

We make the usual assumption that all the galaxies in our study are oblate and none are prolate. We suppose a Gaussian distribution for $\varepsilon_{\rm true}$ (see the top-left plot of Fig. 2). The Gaussian characteristics are chosen so that the simulated $\varepsilon_{\rm app}$ distribution reproduces the observed one. This results in a Gaussian centred at 0.55 with a full width of 0.07. The projection correction depends on the relation between $M/L$ and $\varepsilon_{\rm true}$. The top-right plot of Fig. 2 displays the assumed relation. The simplest assumption is a linear relation. We can also use a Bose–Einstein function to avoid possible unphysical negative $M/L$ near $\varepsilon_{\rm true} = 0$. We use in the linear case the form $\alpha x + \beta$ with $\alpha = 49$ and $\beta = -11$. For the Bose–Einstein function, we use $a/({\rm e}^{(-\frac{x-b}{c})} + 1) + d$ with $a = 45$, $b = 0.5$, $c = 0.1$ and $d = 0$. In both cases, these values are chosen so that the fit of $M/L$ versus $\varepsilon_{\rm app}$ (bottom-right plot) reproduces the experimental results. The bottom-left plot displays the observed (dashed line histogram) and the simulated (plain line histogram) apparent ellipticity distributions. The simulated result is obtained from the Gaussian distribution transformed using equation (1) in which $i$ is random. (We also add a random shift towards smaller $\varepsilon_{\rm app}$ values to simulate the rounding effect of detector resolution.) The bottom-right plot shows the $M/L$ ratio versus $\varepsilon_{\rm app}$. We use all the simulated data to perform the fit, but for clarity, we only show the error bars for the first 100 points. The pattern matches well with what we observe (shown by bigger square symbols) and confirms that the $M/L$ dispersion is not just statistical but mostly due to the random projections of the galaxies. The fitted slope compared to the initial function's slope provides the projection correction. For practical purpose, given the distribution of $\varepsilon_{\rm true}$ (top-left plot), we can restrict ourself to the region $0.35 < \varepsilon_{\rm true} < 0.65$. Averaging the results obtained using the linear or the Bose–Einstein functions, the model simulation indicates a slope correction due to projection effects of about $5 \pm 1$ for the particular results from Bacon et al. (1985). For our global averaged result, we must apply this procedure to $M/L$ ratios extracted using $\varepsilon_{\rm app}$ before they are combined to ratios extracted using $\varepsilon_{\rm true}$. All in all, $\frac{{\rm d}(M/L)}{{\rm d}\varepsilon}$, the slope of $M/L$ with $\varepsilon$, increases by a factor $1.9 \pm 0.3$. The average correction is not approximately five times larger as for the particular case of Bacon et al. (1985) because (1) $M/L$ ratios extracted using $\varepsilon_{\rm true}$ have a 0 correction and (2) the additional uncertainty from the projection correction is proportional to the slope, so the contribution to points near $\langle \frac{{\rm d}(M/L)}{{\rm d}\varepsilon} \rangle \sim 0$ is smaller and their relative weight in the averaging procedure increases.

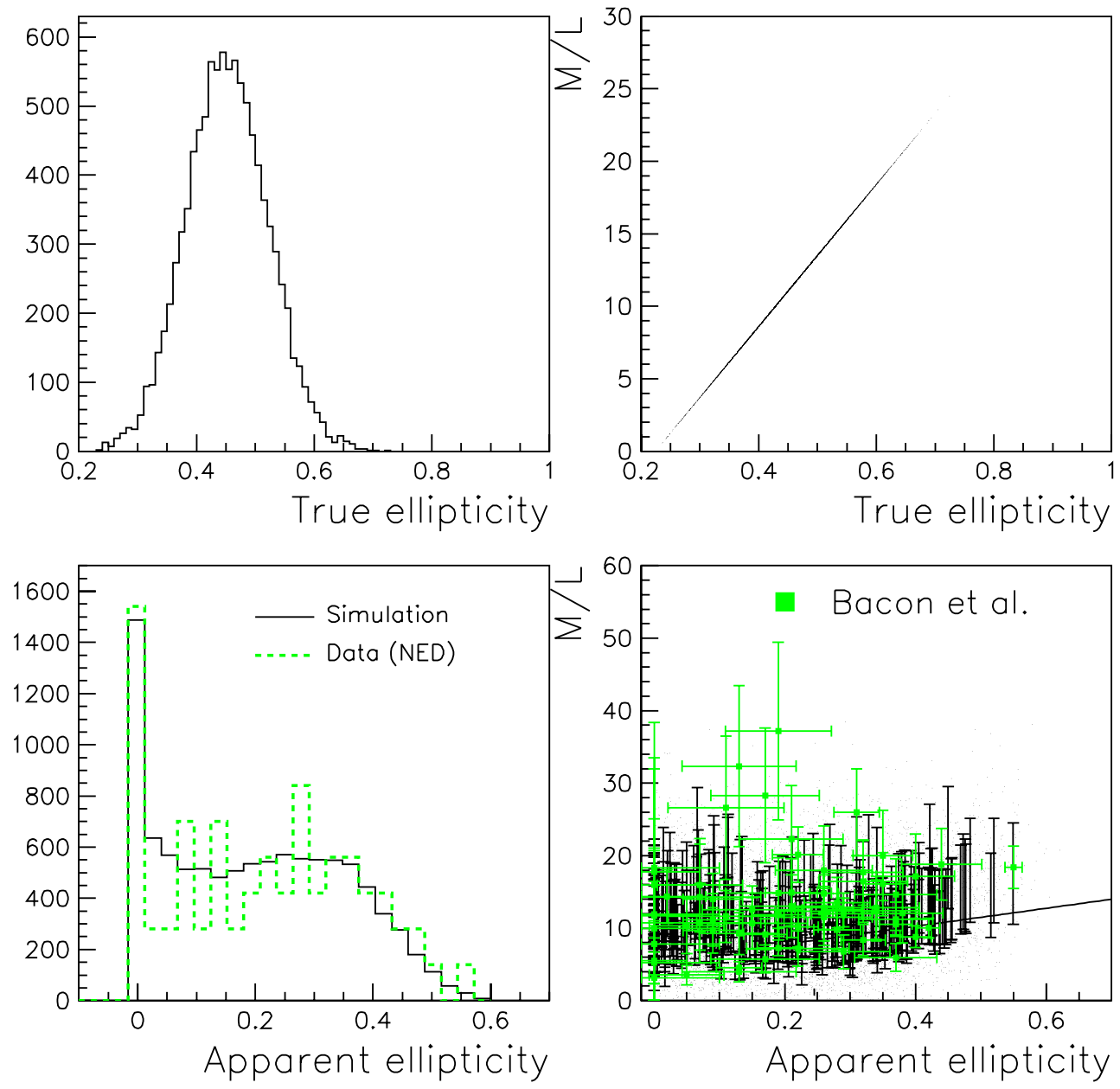

**Figure 2.** Effect of the projection of the 3D elliptical galaxies in our observation plan. Top-left plot: simulated distribution of the true ellipticity $\varepsilon_{\rm true}$. Top-right plot: assumed relation between $M/L$ and $\varepsilon_{\rm true}$ (linear case). Bottom-left plot: apparent ellipticity $\varepsilon_{\rm app}$ (plain line: simulation, dash line: data). Bottom-right plot: $M/L$ versus apparent $\varepsilon_{\rm app}$ (dot: simulation, square: experimental data).

## 7 GLOBAL RESULTS

### 7.1 Normalization

Combining all results requires some care. Most but not all $M/L$ are extracted in the $B$ band. There are scale factors between $B$ band results and those extracted in other frequency bands. Other scaling factors occur if authors used different Hubble constant values. In addition, the $M/L$ ratios are provided at different radii (typically from $0.1R_{\rm eff}$ to a few $R_{\rm eff}$), and it has been shown in many studies (e.g. van der Marel 1991; Capaccioli et al. 1992; Bertola et al. 1993; Kronawitter 2000; Magorrian & Ballantyne 2001; Napolitano et al. 2005; Nagino & Matsushita 2009) that $M/L$ is radially dependent. Finally, sometimes $M_{\rm tot}/M_*$ is provided ($M_*$ is the stellar mass) rather than $M/L$ (see e.g. the top-right plot of Fig. 1), but $M_{\rm tot}/M_*$ is approximately proportional to $M/L$ in most studies. To address these points, we will first normalize to $M/L(\varepsilon_{\rm app} = 0.3) \equiv 8\,{\rm M}_\odot/{\rm L}_\odot \equiv 4M_{\rm tot}/M_*(\varepsilon_{\rm app} = 0.3)$. The values $M/L = 8\,{\rm M}_\odot/{\rm L}_\odot$ and $M_*/L = 4\,{\rm M}_\odot/{\rm L}_\odot$ are typical in the $B$ band for elliptical galaxies. We normalize at $\varepsilon_{\rm app} = 0.3$ because the observed $\varepsilon_{\rm app}$ distribution peaks there (see the dashed line distribution on the lower-left panel of Fig. 2. We should ignore the $\varepsilon_{\rm app} = 0$ peak due the rounding effect of the detector resolution) and because it avoids uncertain extrapolations which would be needed if, for example, we were to normalize at $\varepsilon_{\rm app} = 0$. After discussing first the results obtained with this simple normalization, we will finally use a normalization accounting for the radial dependence of $M/L$. This will be described in Section 7.4.

### 7.2 Corrections for correlations between results

Another point of caution is that some publications use similar methods to extract $M/L$ and these have quoted results for common galaxies. For example, the galaxy NGC 7619 is used in four of the eight publications employing the virial theorem. Consequently, their results may be correlated and may bias the average. In addition, some of the $M/L$ extractions are, in our context, more reliable than others. To address this, the uncertainties (listed in Table A1) are multiplied by the reliability factor given in the same table (rg). This prescription allows some subjectivity but happens to have small influence because the less reliable results usually have also less statistics. We will not discuss how we assigned the reliability factor since this reliability prescription has only a small impact on the global result (not accounting for it, i.e. setting rg = 1 in Table A1, would increase the global result by 2 per cent). It would also necessitate a detailed discussion of the treatment of each paper, which is too long to be exposed here. These details will be given in an archival paper. To correct for the correlation between $M/L$ extractions using similar methods and same galaxies, we count the common galaxies and increase each uncertainty accordingly assuming that they are statistically dominated. (In doing so, we must account for the reliability factors since results are weighted by them when combined.) These weighting factors are listed in Table A1 as well (wf). This is a multiplicative factor to the uncertainties: the higher wf, the less impact the data set has on the global average.

This procedure assumes that results using a given method and an identical galaxy sample are perfectly correlated, which is not exact since within similar methods different assumptions, profiles, input distributions, etc., are used. The specific analyses also differ in details and the data quality varies as the publications used span a range of 30 yr. Thus, the results are not perfectly correlated and consequently, our procedure overestimates the uncertainties. This is partly mended by fitting $\frac{{\rm d}(M/L)}{{\rm d}\varepsilon}$ versus radius, shown on Fig. 3 or 4, with a one-parameter function (i.e. $\frac{{\rm d}(M/L)}{{\rm d}\varepsilon}$ is taken to be constant with radius) and scaling the $\frac{{\rm d}(M/L)}{{\rm d}\varepsilon}$ uncertainties according to the *unbiased estimate* prescription. This is justified because the $\frac{{\rm d}(M/L)}{{\rm d}\varepsilon}$ are from either different methods (for which we can assume that their systematic uncertainties are uncorrelated) or for results using the same method, because we have accounted for the correlation. Although accounting for common galaxies and different reliability is necessary to average as accurately as possible, the net effect happens to be moderate: it increases the slope $\frac{{\rm d}(M/L)}{{\rm d}\varepsilon}$ by a factor 1.11, with the slopes before and after this correction compatible within uncertainties.

### 7.3 Choices of results within a publication

Many publications give $M/L$ extracted with different inputs, such as e.g. different initial mass functions (IMF).

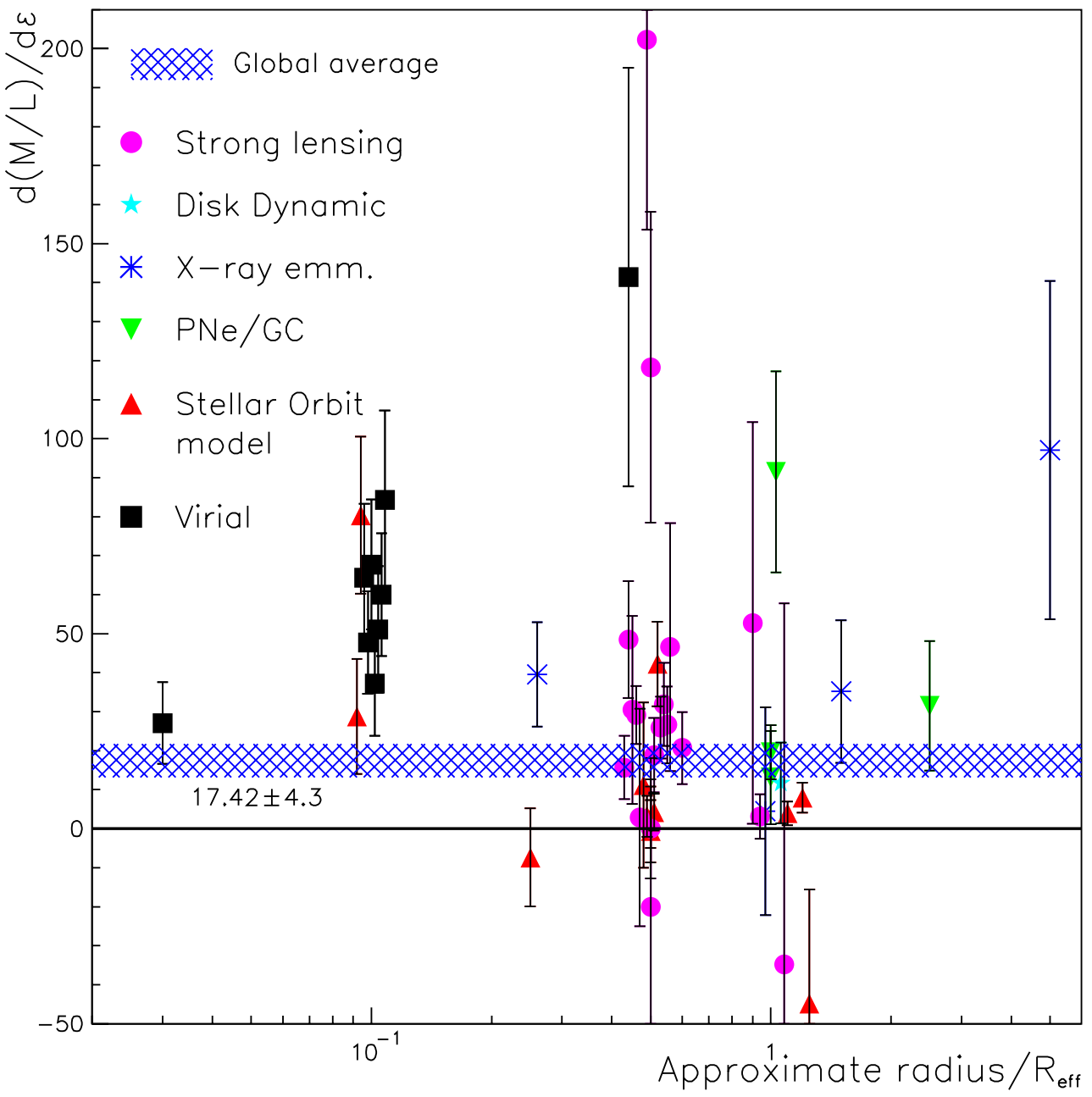

**Figure 3.** Slopes $\frac{d(M/L)}{d\varepsilon_{true}}$ versus approximate radii (in units of $R_{eff}$) at which the $M/L$ are extracted. These results are obtained with the constant normalization $M/L(\varepsilon_{app}=0.3)=8\,M_{\odot}/L_{\odot}$ (see Section 7.1). The six symbol types distinguish the methods used to obtain $M/L$. The band indicates the weighted average value with its uncertainty (given below the band) after accounting for correlations.

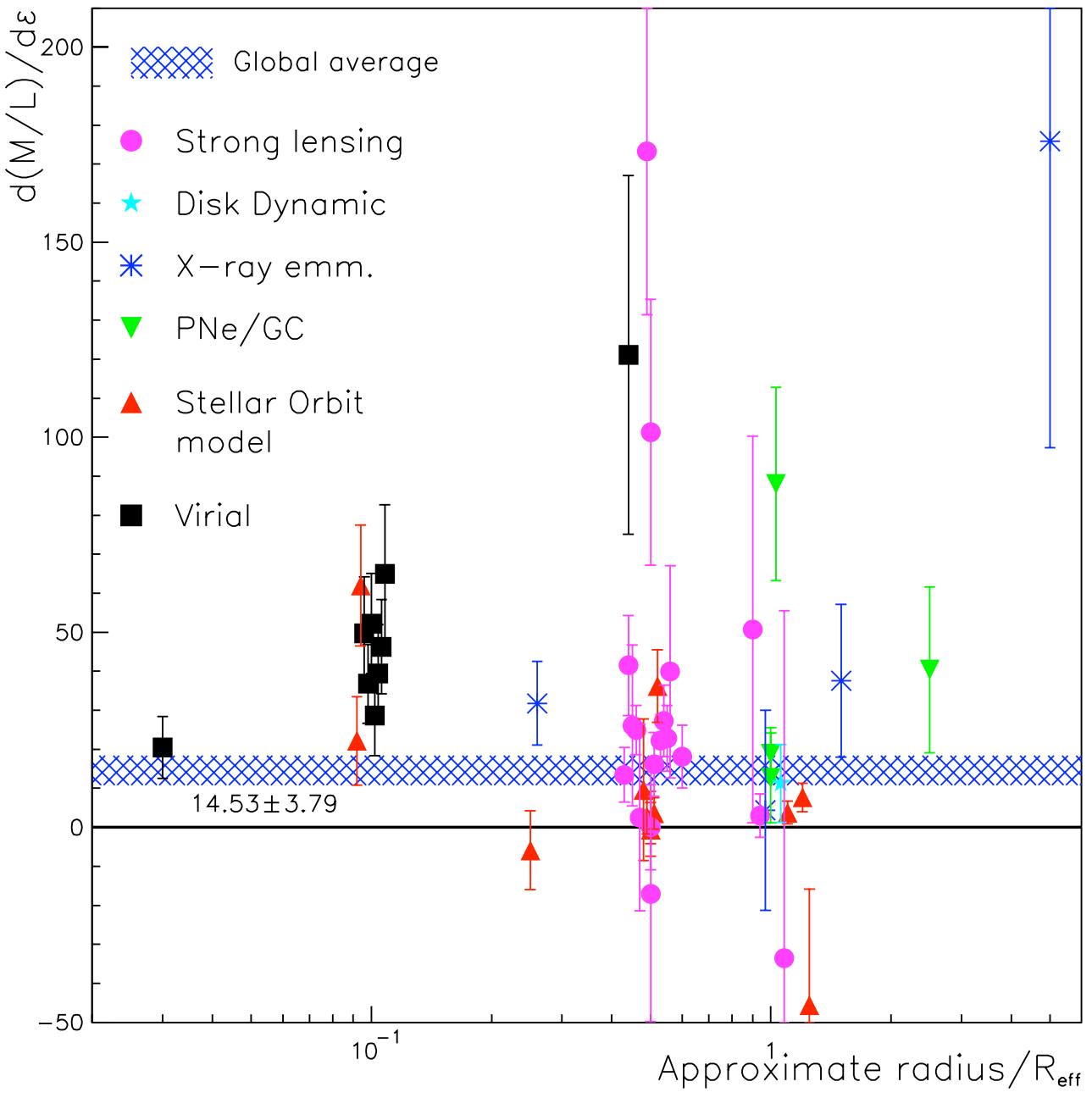

**Figure 4.** Same as Fig. 3, but with the radius dependence of $M/L$ accounted for by using the normalization given by equation (2).

When results are available with several IMF, we chose the Chabrier IMF (Chabrier 2003) since it is most widely used, along with the Salpeter IMF (Salpeter 1955), and is more modern than the Salpeter one.

For $M/L$ extracted using strong lensing, $\frac{d(M/L)}{d\varepsilon}$ can be computed using either $\varepsilon_{app}$ or mass ellipticity $\varepsilon_{mass}$. The latter is the ellipticity of the modelled total mass distribution. We chose to use $\varepsilon_{app}$, as it is used for the other five methods. We note however that 8 out of the 10 strong lensing publications for which $\varepsilon_{mass}$ is available display a stronger $M/L$ versus $\varepsilon_{mass}$ correlation, i.e. $\frac{d(M/L)}{d\varepsilon_{app}} < \frac{d(M/L)}{d\varepsilon_{mass}}$.

### 7.4 Normalization accounting for the radius dependence of $M/L$

The slopes $\frac{d(M/L)}{d\varepsilon_{true}}$ computed with the constant normalization $M/L(\varepsilon_{app}=0.3)=8\,M_{\odot}/L_{\odot}$ are shown in Fig. 3. They are plotted for clarity against the approximate average radii at which the $M/L$ values have been extracted. (Some papers provide $M/L$ calculated at different radii, so several points may correspond to one publication. Also, a few outlying points are outside the graph range. Consequently, there are more than 41 points in Fig. 3. We have a total of 50 data points.) One notices systematic shifts between results using different methods, most noticeably between the strong lensing and virial results. This may be due to the different radii at which the $M/L$ values are extracted: as investigated e.g. in Bertola et al. (1993), Capaccioli et al. (1992), Kronawitter (2000), Magorrian & Ballantyne (2001), Nagino & Matsushita (2009), Napolitano et al. (2005), Thomas et al. (2007, 2011) and van der Marel (1991), baryonic matter appears to dominate over dark matter at smaller radii. Thus, $M/L$ increases with radius. However, we normalize the $M/L$ to the common $8\,M_{\odot}/L_{\odot}$ value, effectively taking out this dependence. Hence, by assuming that $\frac{d(M/L)}{d\varepsilon}$ is originally independent of radius, we introduce a spurious dependence with radius since $\frac{d(M/L)}{d\varepsilon}$ is proportional to the factor used to scale $M/L$ to the normalization value. This spurious dependence is opposite to the $M/L$ dependence with radius, i.e. $\frac{d(M/L)}{d\varepsilon}$ decreases with increasing radius, as seen by the general trend on Fig. 3. To avoid this problem, we account for the dependence of $M/L$ with radius $r$ by using the normalization

$$M/L(\varepsilon_{app}=0.3)=(6+1.7r/R_{eff})\,M_{\odot}/L_{\odot}. \quad (2)$$

This dependence is suggested by the study in Capaccioli et al. (1992). The slopes $\frac{d(M/L)}{d\varepsilon}$ obtained with this $r$-dependent normalization are shown in Fig. 4. Comparing Figs 3 and 4, we see that the radius dependence of $\frac{d(M/L)}{d\varepsilon}$ is reduced (the slope of the general trend in Fig. 4 is smaller than the one in Fig. 3). To fully cancel it, we would need a factor 3–4 rather than 1.7 in equation (2). Such stronger dependence would still be compatible with the results given in Capaccioli et al. (1992). Hence, the radius dependence seen in Fig. 3 may be fully explained by the radius dependence of $M/L$. However, other factors may contribute to the systematic shifts seen in Fig. 3: a looser selection is applied on the distant – and thus less characterized – strong lensing galaxies, so contamination may dilute the correlation. (We will show in Section 8.4 that indeed the looser selection for distant galaxies tends to reduce the correlation.) In addition, no ellipticity correction is applied to the $M/L$ extracted using strong lensing because this method should be less sensitive to $\varepsilon$. Interestingly, if we were to consider $\frac{d(M/L)}{d\varepsilon_{mass}}$ rather than $\frac{d(M/L)}{d\varepsilon_{app}}$, the strong lensing and the corrected virial results would fall in line. Finally, strong lensing galaxies tend to be more massive than the local ones, which may induce a systematic difference.

### 7.5 Results

As seen in Fig. 4, all the averaged $\frac{d(M/L)}{d\varepsilon_{true}}$ from the six different methods are non-zero and positive. They are listed in Table 1. Combining all data sets leads to $\langle\frac{d(M/L)}{d\varepsilon}\rangle=(14.53\pm3.79)\,M_{\odot}/L_{\odot}$. With our initial constant normalization $M/L(\varepsilon_{app} = 0.3) = 8\,M_\odot/L_\odot$, i.e. averaging the results shown on Fig. 3, we would have obtained $\langle\frac{d(M/L)}{d\varepsilon}\rangle = (17.42 \pm 4.30)\,M_\odot/L_\odot$, We conservatively account for the uncertainty associated with the $r$-dependence of the normalization by taking the difference between the two results obtained with the constant and the $r$-dependent normalization. We then finally obtain,

$$\left\langle \frac{d(M/L)}{d\varepsilon} \right\rangle = (14.53 \pm 4.77)\,M_\odot/L_\odot, \tag{3}$$

a statistically meaningful positive signal, indicating that $M/L$ and $\varepsilon$ correlate. We note that, without the selection criteria, the otherwise clear correlation is obscured by large fluctuations, which may explain why it has not been noticed earlier. For example the recent detailed analyses (Cappellari 2013a,b) found no sign of such correlation but, while their galaxy sample is large (260 galaxies), only about 10 per cent of them are suitable elliptical galaxies and about 80 per cent of the galaxies are spiral and lenticular galaxies, which are excluded by our selection criteria in Section 3. We also verified with the data (Bacon et al. 1985) that the correlation disappears when the selection is removed.

**Table 1.** $d(M/L)/(d\varepsilon)$ for the six methods used to extract $M/L$.

| Method | $\frac{d(M/L)}{d\varepsilon}$ |
|---|---|
| Virial | $46.77 \pm 16.00$ |
| Stellar orbit | $3.92 \pm 5.74$ |
| PNe/GC | $29.22 \pm 19.63$ |
| X-ray | $38.21 \pm 12.29$ |
| Disc Dyn. | $9.38 \pm 14.52$ |
| Strong lensing | $15.72 \pm 6.84$ |

The value of $\frac{d(M/L)}{d\varepsilon}$ scales with our normalization $M/L(r = R_{eff}, \varepsilon_{app} = 0.3) = 7.7\,M_\odot/L_\odot$, see equation (2). Compared to $7.7\,M_\odot/L_\odot$, the slope is large.

One may suspect a large influence of the virial data with their large number of data sets and small uncertainties but which are extracted at small radii, i.e. inside the region where dark matter is not dominant. However, excluding all virial data leads to a similar correlation: $\langle\frac{d(M/L)}{d\varepsilon}\rangle = (12.61 \pm 3.90)\,M_\odot/L_\odot$. The virial data have a relatively small influence because (1) these data are correlated (similar sets of galaxies, same method) and so are weighted out by the procedure described in Section 7.2 (see Table A1), and (2) these older data tend to be assigned a lower reliability than newer data using more sophisticated extraction methods, see Table A1. (These two factors lowering the influence of the virial data are not apparent on Figs 3 and 4.)

We also performed a survival analysis in which each data set is removed in turn and the correlation re-estimated. The resulting distribution of the $d(M/L)/d\varepsilon$ has a 0.514 rms (the average $d(M/L)/d\varepsilon$ value is 14.54). This is to be compared with our quoted uncertainty of 4.77 (for a $d(M/L)/d\varepsilon$ value of 14.53). The rms is significantly smaller than our quoted uncertainty and is statistically less rigorous since the data sets do not have the same weights when averaged, and some are partially correlated. However, this demonstrates that the correlation reported here is not due to a bias from a specific data set.

### 7.6 Quality of the uncertainty estimate

Our signal has a $3\sigma$ significance. One may wonder what would be the influence of underestimated uncertainties in individual data sets on the significance of our result, since it is known that sometimes reported uncertainties are underestimated. To be certain that our reported correlation does not stem from underestimated uncertainties that are giving a false significance to a statistical fluctuation, we used, as already discussed, the unbiased estimate procedure to reassess the uncertainties in each data sets. Hence, the uncertainties in each data set represent correctly the point to point scatter of the $M/L$ data, assuming that it is random. After combining the individual results, we applied again the unbiased estimate procedure to the obtained $d(M/L)/d\varepsilon$ data set. This accounts for possible systematic biases between the data sets used to extract $M/L$. This effectively increases the uncertainty on each data sets, as seen in Section 7.5 when we compared the rms from the survival analysis (0.514) to our final uncertainty (4.77).

Our final uncertainty estimate should thus encompass most of the statistical and systematical uncertainties. However, it does not account for possible systematic effects affecting all data sets and all techniques. We investigate such possibility in the next section.

## 8 SYSTEMATIC STUDIES

In this section, we summarize the systematic tests conducted to check whether biases are at the origin or contribute to the non-zero value of $\frac{d(M/L)}{d\varepsilon}$.

### 8.1 Methodological bias

The first systematic bias to expect is a methodological bias from a model or an extraction method that may be systematically not describing well some aspect of the galaxies. However, as already discussed, this is likely ruled out because the galactic dark matter content is extracted with six very different approaches and, to smaller extent, because we employed many publications that generally used related but different models.

### 8.2 Bias in the correlation assessment

We assess the existence of correlation by fitting data with a linear fit. This assumes a linear dependence of $M/L$ with $\varepsilon$. We used the data from Bacon et al. (1985) to check whether a second-order polynomial would describe better the data. However, it yields a $\chi^2/ndf$ similar to the linear fit (for this comparison, we used the uncertainties provided by the authors and did not rescale them).

Alternatively, to assess the correlation, one can compute the Pearson correlation coefficient $p$ given by the covariance of $\varepsilon$ and $M/L$ divided by their standard deviations: $p = \frac{\mathrm{cov}(\varepsilon, M/L)}{\sigma_\varepsilon \sigma_{M/L}}$, see e.g. Spiegel (1992). We have $|p| \leq 1$ and larger values of $|p|$ indicate clearer (smaller dispersion) and/or stronger (steeper slope) correlations. However, since this type of statistical analysis does not account for individual uncertainties, contrarily to a $\chi^2$ minimization fit, it is ill-suited for our samples that display a large range in uncertainties. This problem can be partly circumvented by keeping data of a given absolute precision. We applied this procedure to the data (Bacon et al. 1985), keeping points for which the uncertainty on $M/L$ is smaller than $5\,M_\odot/L_\odot$. In that case, the sample is reduced to 48 galaxies. We first checked the value of $p$ before applying the projection correction discussed in Section 6. We found $p = 0.367$. Applying a tighter selection $\triangle M/L < 3\,M_\odot/L_\odot$ reduced the sample to 23 galaxies and yielded $p = 0.511$. Such values of $p$ reveal a medium to large correlation between $M/L$ and $\varepsilon$ and confirm the conclusion from the fit method. Selecting the highest precision data strengthen the $M/L$ versus $\varepsilon$ correlation (both for the determination using of the Pearson criterion and for the determination from the linear fit). After applying the projection correction described in Section 6, the Pearson correlation coefficients are 0.87 and 0.91 after removing the galaxies with (unscaled) uncertainty above $\triangle M/L = 5$ and $\triangle M/L = 3$, respectively. In both cases, they indicate a clear and strong correlation. Again, it is interesting to notice that if we select the highest precision data, then the correlation is enhanced, both for the determination using the Pearson criterion and for the determination from the linear fit.

From these studies, we conclude that the observed non-zero value of $\frac{\mathrm{d}(M/L)}{\mathrm{d}\varepsilon}$ is not due to an inappropriate assessment of the $M/L$ versus $\varepsilon$ correlation.

### 8.3 Effects of relaxed selections

In some instances, we relaxed our selection criteria to have enough galaxies for a meaningful study. In particular, we sometimes relaxed the distant galaxy criterion $\sigma \leq 225\,\mathrm{km\,s^{-1}}$ that should minimize the contamination from S0. However, such selection may also remove genuine elliptical galaxies, possibly biasing the results. A large sample of well-identified elliptical galaxies provides the opportunity to check the effect of removing from the sample suitable elliptical galaxies with low velocity dispersion ($\sigma < 225\,\mathrm{km\,s^{-1}}$). Such a check was done using the set of 102 suitable elliptical galaxies from Prugniel & Simien (1996). Once the $\sigma > 225\,\mathrm{km\,s^{-1}}$ selection is applied, only 35 galaxies remain. The mean value of $M/L$ increases since $M/L \propto \sigma$. However, normalizing the fit result to the same average $\langle M/L \rangle$ value obtained in our main study of Prugniel & Simien (1996), we get $M/L_B = (7.10 \pm 1.05)\varepsilon_{\mathrm{app}} + (2.47 \pm 0.21)$, in good agreement with the main result $M/L_B = (6.58 \pm 1.98)\varepsilon_{\mathrm{app}} + (2.41 \pm 0.41)$ (see Table A1 in the appendix). This indicates that no noticeable bias arises from the $\sigma > 225\,\mathrm{km\,s^{-1}}$ requirement. This test is statistically significant since more than 50 per cent of the 112 galaxies have been removed.

In other instances we relaxed our selection to include LINER galaxies. Using again the data set from Prugniel & Simien (1996), we repeated the linear fit using only LINER galaxies. It is compatible with the main result, albeit with larger statistical uncertainty.

Using the data set from Bacon et al. (1985), we also checked the effect of removing from the sample two galaxies that are close to the dwarf elliptical galaxy locus (we imposed $M_B > -17.8$). This had no significant consequence on the $M/L$ versus $\varepsilon$ relation.

Finally, we investigated the effect of magnitude/mass/shape selection on the correlation. Using data from Auger et al. (2010) and the luminosities from Bolton et al. (2008), we grouped the galaxies into two sets, one of 25 galaxies with magnitude $M_B \geq -19.5$ (this selects less massive discy galaxies) and the other containing 17 galaxies with magnitude $M_B \leq -19.5$ and no generic selection $M > 10^{12}\,\mathrm{M_\odot}$ on the total mass (this selects more massive boxy galaxies). We found no clear difference between samples of luminous/boxy galaxies or of fainter/discy galaxies.

### 8.4 Environment effects

In general, our local galaxy selection criteria should ensure that the selected galaxies are not or were not recently significantly influenced by their neighbours. Such assurance cannot be warranted for distant galaxies. Those are used mostly with the strong lensing method for which, although equilibrium is not required, internal or external shears due to interaction with nearby neighbours can bias the determination of the mass. Shear effects are usually corrected by the authors, although such corrections are not unambiguous. Lensing galaxies are often found in clusters or groups. We studied the effects of environment using the data sets (Cardone et al. 2009, 2011; Auger et al. 2010; Barnabè et al. 2011). We compared $\frac{\mathrm{d}(M/L)}{\mathrm{d}\varepsilon}$ obtained for field galaxies with those obtained for galaxies found in groups or clusters, according to Treu et al. (2009). The analysis of sets (Auger et al. 2010; Barnabè et al. 2011; Cardone et al. 2011) shows that the galaxies in groups or clusters display more dispersion in their $M/L$ versus $\varepsilon$ correlation. [We found an opposite result with Cardone et al. (2009) but it is compatible with a statistical fluctuation.] Such a result is expected if unaccounted galaxy–galaxy interactions cause data jitter. In addition, $\frac{\mathrm{d}(M/L)}{\mathrm{d}\varepsilon}$ is larger for isolated galaxies. In order to keep enough statistics, we still used indiscriminately field and group/cluster galaxies in our analyses, even if this may dilute the correlation.

### 8.5 Anisotropy effects in the virial method

In the virial method, unphysical correlations between $M/L$ and $\varepsilon$ can be induced by anisotropic star motion. However, as assessed in Bacon et al. (1985), this effect is small. Furthermore, anisotropies have little effects on the slopes of figs 1a and b of Bacon et al. (1985) while affecting primarily the $M/L$ absolute values. Since we are investigating a dependence of $M/L$ on $\varepsilon$ and not primarily the absolute scale of $M/L$, possible effects of the anisotropies are not critical. (In any case, the values of $M/L$ are normalized with equation 2 when all the data sets are combined.) In addition, bright galaxies – which tend to display these anisotropies – are discarded from our samples by our selection criteria. Finally, as will be discussed in Section 8.6, we checked that the results from Bacon et al. (1985), which used a large galaxy set and the virial method, display no significant $M/L$ versus brightness correlation, which is an a posteriori justification.

### 8.6 Correlations

There are many interrelated quantities describing galaxies. They may induce spurious correlations with $M/L$ or $\varepsilon$: a possible measurement bias or observation bias not apparently related to $M/L$ and $\varepsilon$ can propagate to them via correlations. We mostly used the data set from Bacon et al. (1985) to check such possibilities. Using this set is convenient because it has large statistics and provides nine galactic characteristics to check. Those are as follows.

(i) $M/L$ (from three estimates, see Table A1).
(ii) Absolute effective radius $Re$(Kpc).
(iii) Apparent effective radius $Re$(arcsec).
(iv) Central velocity dispersion $\sigma_0$.
(v) Absolute blue magnitude $M_B$.
(vi) Distance modulus $DM$.
(vii) Integrated apparent blue magnitude $B_t$.
(viii) Surface brightness $I_e$.
(ix) Apparent axis ratio $R_m/R_M = 1 - \epsilon_{\mathrm{app}}$.

We used also other sets for checking galactic characteristics not provided in Bacon et al. (1985).

The data of Bacon et al. (1985) are old but since we selected galaxies for which characteristics agree with the up-to-date NED ones, this is not a problem. In any case, checking for effects of biases with old data, presumably more prone to measurement biases,

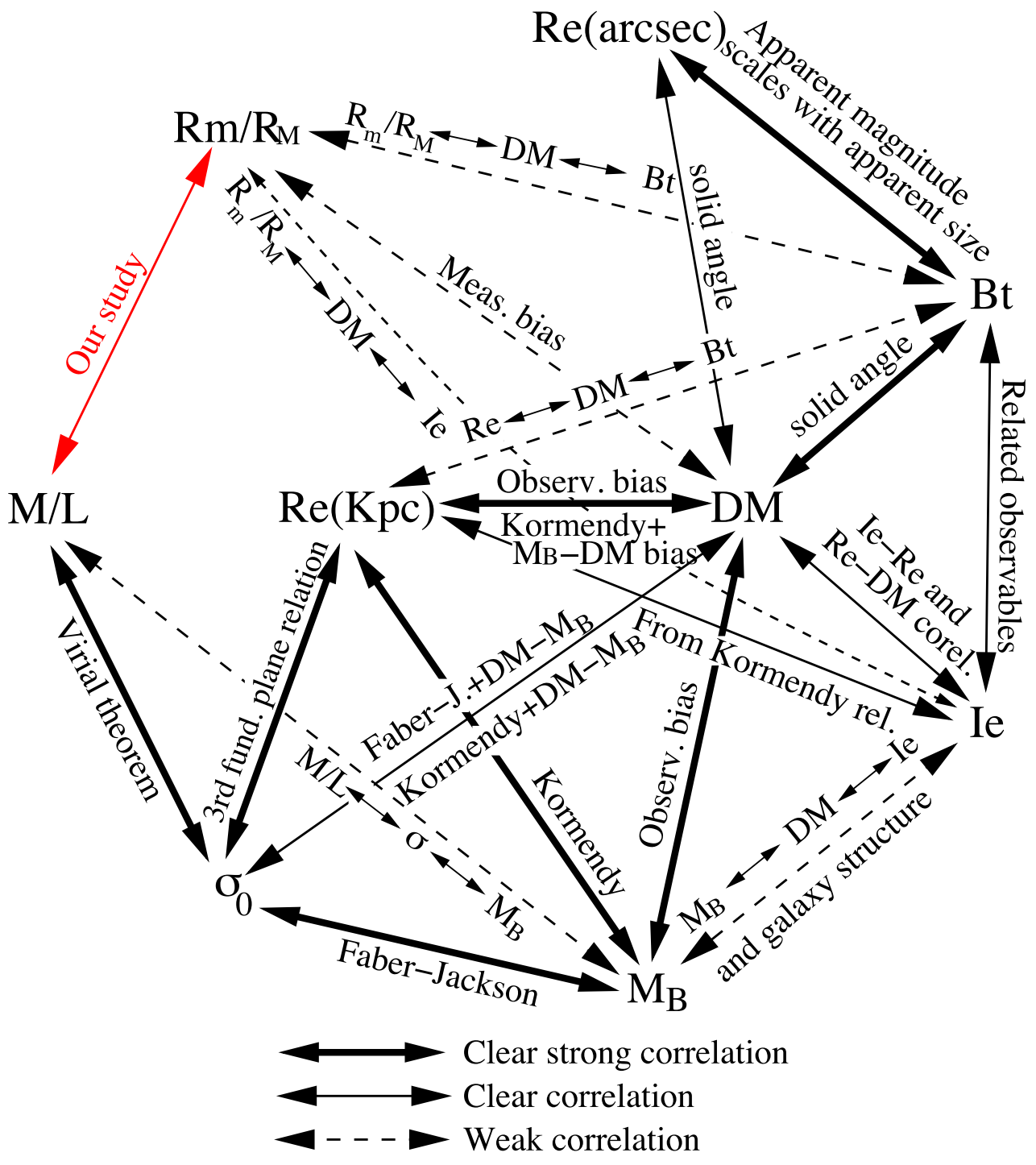

**Figure 5.** Observed correlations for the data set (Bacon et al. 1985). The thick arrows indicate clear strong correlations (based on a clearly non-zero slope of linear fits), the thin arrows indicate clear correlations and the dashed arrows indicate weaker correlations. The lighter arrow is the correlation we study. The origin and name of the correlations are written near the corresponding arrows.

should give an upper limit on the effect of these biases. There are three types of correlations to expect:

(1) Known trivial relations, e.g. the decrease of apparent intensities with the galactical distance to Earth, or the relation provided by the virial theorem between $\sigma$ and $M/L$.

(2) Phenomenologically known correlations, e.g. the Faber–Jackson (Faber & Jackson 1976) or Kormendy (Kormendy 1977) relations.

(3) Unknown correlations.

We investigated all possible correlations by linearly fitting 2D distributions and checking whether the slopes are compatible with zero. We used the uncertainties provided in Bacon et al. (1985) without rescaling them. We verified whether all expected correlations were seen (types 1 and 2 above). All are clearly seen except for the absolute blue magnitude $M_B$–surface brightness $I_e$ correlation. Such a correlation was reported in Binggeli (1994). We do see a correlation but not as clearly as the other expected correlations. In addition, we found a number of other, weaker, correlations that are either the consequences of expected correlations or may be due to measurement biases. All those results are summarized in Fig. 5. Without our selection criteria, we would have expected other correlations than those shown in Fig. 5, for example between *Re*(Kpc) and $R_m/R_M$ since small galaxies tend to be more elongated, or between *Re*(Kpc) and $M/L$ (dwarf and giant galaxies tend to have larger $M/L$). Also, dwarf and giant galaxies tend to be rounder leading to an $M/L$–$R_m/R_M$ correlation. These correlations would have clouded our study and this supports our rejection of these classes of galaxies.

We see on Fig. 5 that there is no direct correlations between both $M/L$ and $R_m/R_M$ and a third quantity. This excludes, at least for the data set (Bacon et al. 1985), the possibility that the $M/L$ and $\varepsilon = 1 - R_m/R_M$ correlation is a spurious effect of correlations between some of the other seven quantities in Fig. 5. However, the other correlations can still weakly contribute to the $M/L$–$\varepsilon$ correlation in a multistep manner. For example, $M/L$ correlates with $\sigma_0$, which correlates with the distance modulus, which correlates (weakly) with $R_m/R_M$. Such possibilities are discussed in Section 8.7.

In addition to the correlations between quantities listed in Bacon et al. (1985), we also used the data from Lauer (1985) to check the influence of $M/L$ correlating with metallicity. Lauer noticed a correlation between galaxy metallicity $M_{g_2}$ and $M/L$, which we verified. However, we found no clear correlation between $M_{g_2}$ and $\varepsilon$. Consequently, the correlation between $M_{g_2}$ and $M/L$ should not contribute to $\frac{d(M/L)}{d\varepsilon}$. In addition, Lauer signalled a correlation between the luminosity density $\rho$ and $M/L$. Again, we verified this but found no clear correlation between $\rho$ and $\varepsilon$.

While we attempted to be thorough in our search for systematic biases, all possibilities cannot be excluded. For example, the role of dust has not been investigated because elliptical galaxies are typically considered to be dust-free. However, it was recently noticed that 6 per cent of them show signs of dust (Rowlands et al. 2012). The small additional mass it represents is insignificant but most likely, extinction from dust varies with inclination. Considering the simplest case of a thin dust disc in the equatorial plane of an oblate galaxy, dense enough dust would cause a larger extinction for galaxies viewed face-on, leading to an apparent increase of $M/L$ ratios at small $\varepsilon$ (a trend opposite to the one reported here). Although this most naive expectation would suppress the trend observed rather than create it, this illustrates how an overlooked attribute could explain the observation reported here.

## 8.7 Influence of measurement biases

As seen on Fig. 5, and maybe not surprisingly, the distance modulus *DM* is involved in all the weak correlations involving possible measurement biases. In order to study the effect of *DM* on $\frac{d(M/L)}{d\varepsilon}$, we binned *DM* and plotted $\frac{d(M/L)}{d\varepsilon}$ for each of the *DM* bins. (Again, for this study, we used the data from Bacon et al. 1985.) Except for one outlying point, there is no strong dependence of $\frac{d(M/L)}{d\varepsilon}$ with *DM*: We found $\frac{d(M/L)}{d\varepsilon} = (-3.42 \pm 2.00)DM - 123.5 \pm 63.59$ for a reduced $\chi^2/ndf = 1.8$. This suggests that a correction for the *DM* bias would increase $\frac{d(M/L)}{d\varepsilon}$. The study also suggests that there may be something wrong with the outlying bin (possibly because of its lowest datum, galaxy NGC 4510 that has very small error). Excluding this bin, we find a similar $\frac{d(M/L)}{d\varepsilon} = (-3.25 \pm 2.00)DM - 118.7 \pm 62.94$ for $\chi^2/ndf = 0.8$. The value of $\chi^2/ndf$, closer to the expected 1, supports excluding the bin.

All in all, the correction for a possible *DM* bias would increase $\frac{d(M/L)}{d\varepsilon}$ by a factor $1.6 \pm 0.81$ but this assumes the reliability of a linear extrapolation over 31 units of *DM*, based on a fit performed only over a *DM* range of 4.5 units. We conservatively chose to not apply the correction to $\frac{d(M/L)}{d\varepsilon}$ for the following reasons: (1) the uncertainty attached to the extrapolation is large; (2) the bias may not be real (since the fit coefficient is only 1.6 sigma from zero); (3) ignoring the correction reduces the signature of the correlation we are investigating.

Ignoring the outlying bin from our analysis yields a Pearson coefficient $p = 0.44$ when keeping data with $\Delta M/L < 5$). We find

$p = 0.64$ when keeping data with $\Delta M/L < 3$. The $\frac{\mathrm{d}(M/L)}{\mathrm{d}\varepsilon}$ is 14 per cent greater when the outlying bin is ignored.

These numbers indicate the effect of indirect correlations which, again, may be an upper limit since the data (Bacon et al. 1985) are old. [Conservatively, we do not use these numbers to correct the $M/L$ versus $\varepsilon$ results obtained from Bacon et al. (1985) that we used in the global analysis.]

### 8.8 S0 contamination

It has long been known that S0 and elliptical galaxies can be difficult to distinguish. The distinction becomes harder with decreasing $\varepsilon_{\mathrm{app}}$ (D'Onofrio et al. 1995; Graham et al. 1998). Without additional dynamical information, this difficulty remains with newer data (Bamford et al. 2009; Cappellari et al. 2011, 2013a,b; Emsellem et al. 2011). If the $M/L$ of S0 tend to be smaller than those of elliptical galaxies, an $\varepsilon_{\mathrm{app}}$-dependent S0 contamination of the E sample would induce the observed increase of $M/L$ with $\varepsilon$. In that case, the origin of the correlation would be a systematic bias in classification rather than physical. The tight generic selection criterion discussed in Section 3, minimizes such contamination since we systematically rejected galaxies that are not clearly classified as E, in particular E?, E/S0 or E+ types. In spite of the tight selection, contamination may still be present. We assessed whether the S0 rejection is adequate with the following tests.

**Independent S0 rejection criterion**

An independent second selection criterion added to the primary one can verify if the primary selection is inefficient and the effect of a resulting contamination. Such additional criterion was used in Section 8.3. Adding the second criterion yields a result compatible with the nominal analysis and this validates in principle our primary S0 rejection. One possible caveat is that the second selection could add a new bias. If an effect from S0 contamination were present in the nominal analysis, to explain the agreement between results with and without second selection, one must invoke a new bias introduced by the second selection that would by chance compensate the (now suppressed) effect from S0 contamination. This is unlikely. Another possibility that could explain the agreement between results with and without second selection would be that S0 and E have a similar $M/L$ correlation with $\varepsilon$. In that case, our test would be inefficient to assess the contamination but this one would not affect our final conclusion. In any case, such possibility appears to be ruled out by the test discussed next.

**Estimate of the assumed S0 contamination and consequences**

For this test, we assume that the $M/L$ correlation with $\varepsilon_{\mathrm{app}}$ is due to S0 contamination and check if the consequences are reasonable. To avoid possible systematic bias, we performed the analysis on data obtained using either the virial theorem (Prugniel & Simien data set) or lensing (Auger et al. data set). The studies agree that the assumption leads to results clearly incompatible with the census of S0 and E.

Since the S0 contamination is minimal for large $\varepsilon_{\mathrm{app}}$, $M/L$ ($\varepsilon_{\mathrm{app}} \simeq 0.7$) provides a clean value for $M/L_E$, the $M/L$ for ellipticals, now assumed to be constant with $\varepsilon$. To determine $M/L_{S0}$, we select bona fide S0. For local galaxies, we reject the peculiar S0, dwarf S0, S0?, E/S0, SB0 (since the presence of a bar would betray an S0 even face-on, so they cannot contribute to a contamination of the E galaxy sample), BrClG, S0 belonging to the Arp's catalogue (Arp 1966), etc., similarly to the discussion in Section 3. For the lensing data, we selected galaxies classified as S0 (we rejected the S0/SA and E/S0). From this, we extract a contamination $c(\varepsilon_{\mathrm{app}})$, see e.g. Fig. 6 for the lensing data (the result from the virial data sample is similar). The contamination clearly has to be unreasonably large to explain the observed $M/L$ versus $\epsilon$ correlation. For example, it would imply that essentially all round E galaxies observed are actually misclassified S0. There is uncertainty in our estimate but even conservatively using the $2\sigma$ lower bound of the $c(\varepsilon_{\mathrm{app}})$ band leads to a large $c(0 < \varepsilon_{\mathrm{app}} < 0.1)$: 0.84 (lensing) or 0.54 (virial). Since about 5 per cent of randomly oriented discs would have $\varepsilon_{\mathrm{app}} = 0.05$, and since from the distribution of E with $\varepsilon_{\mathrm{app}}$ (see e.g. Fig. 2) about 30 per cent of the galaxies classified as E have $\varepsilon = 0.05 \pm 0.05$, then using the worst case (virial result, $c = 0.54$), we estimate that S0 are at least seven times more numerous than medium-size E galaxies. This conservative $2\sigma$ estimate is not compatible with the observation that the numbers of S0 and E are similar, see e.g. Calvi et al. (2012), or estimates from stellar mass budget, e.g. Driver et al. (2008), although these estimates can vary by a factor 3, exemplifying the uncertainties in attempting to assess the degree of S0 contamination. Using the lensing results leads to an even greater discrepancy. This implies that the assumption that S0 contamination is at the origin of the correlation is not tenable.

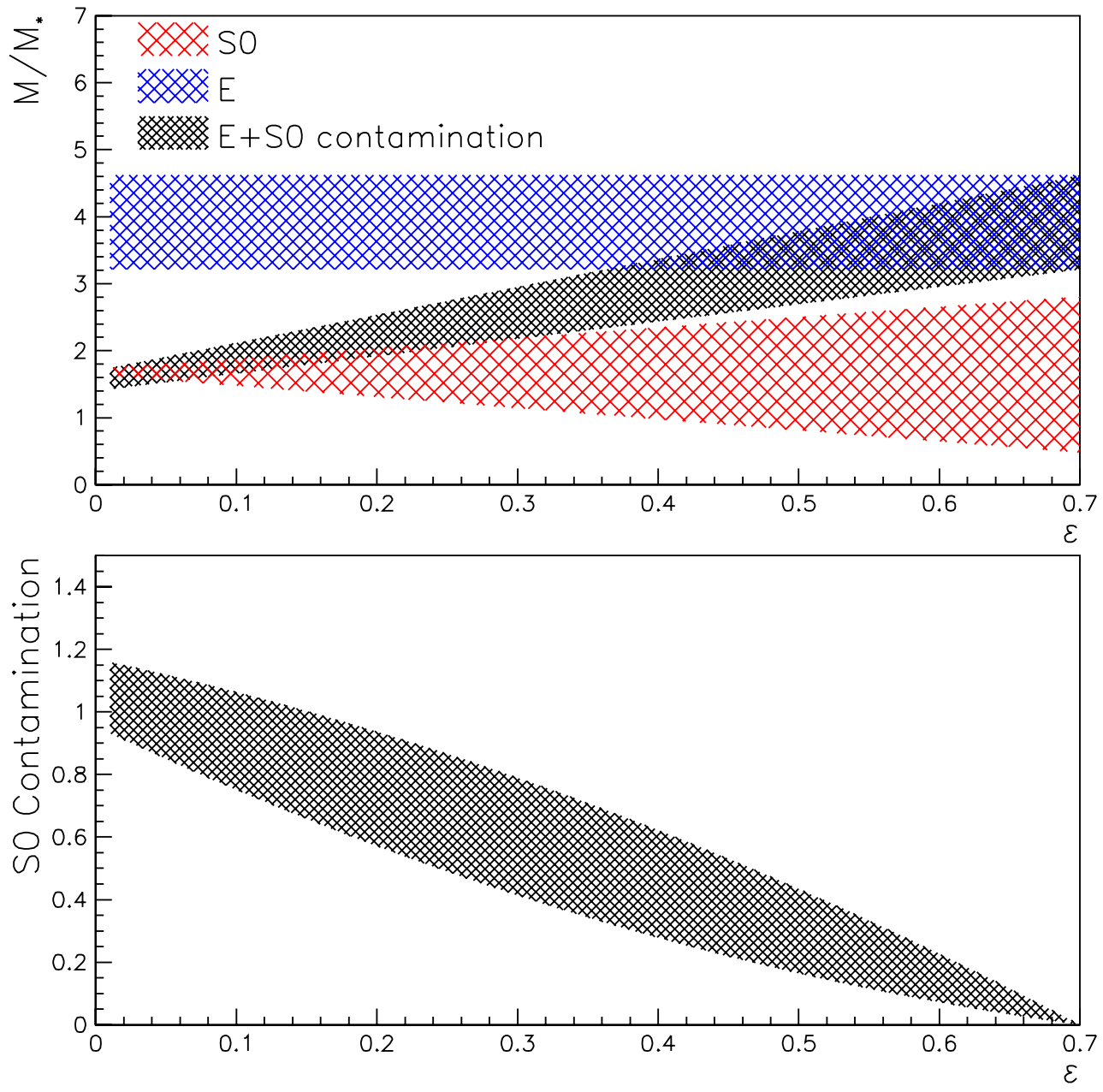

**Figure 6.** Top panel: measured correlation (middle band) and $M/M_{\star S0}$ (lower band), and assumed constant $M/M_{\star E}$ for elliptical galaxies (higher band). Bottom panel: the vertical axis gives $c(\varepsilon)$, the resulting S0 contamination of the E sample assuming that the signal on the top panel is solely caused by S0 contamination.

**Case of S0 contamination at large $\varepsilon$ only**

In the previous section, to explain the linear correlation between $M/L$ and $\varepsilon$ we assumed that the S0 contamination is smoothly varying with $\varepsilon$. However, the linearity could be an artefact of the fit form, see the discussion in Section 8.2. We could instead imagine that a large S0 contamination present only at small $\varepsilon_{\mathrm{app}}$ drives the fit (although if so, higher order forms would fit better the data, which is not the case). If face-on S0 contamination were important, excluding low $\varepsilon_{\mathrm{app}}$ galaxies should markedly reduce $\mathrm{d}(M/L)/\mathrm{d}\varepsilon_{\mathrm{app}}$. We tested this on the data set with highest statistics (Prugniel & Simien, 1996). Removing galaxies with $\varepsilon_{\mathrm{app}} < 0.1$ reduced the statistics from 102 to 76 galaxies. $\mathrm{d}(M/L)/\mathrm{d}\varepsilon_{\mathrm{app}}$ becomes $10.32 \pm 2.89$, to be compared to the nominal value listed in Table A1, $6.58 \pm 1.98$. These numbers are compatible and the new $d(M/L)/d\varepsilon_{app}$ is not smaller than the nominal one, as would have been expected in the case of face-on S0 contamination. One can also see directly on Fig. 1, especially on the top-right panel and the bottom panels, that removing low $\varepsilon_{app}$ galaxies would not affect significantly $d(M/L)/d\varepsilon_{app}$.

**Correlation with distance modulus**

If the correlation was due to S0 contamination, then $d(M/L)/d\varepsilon$ should be larger for distant galaxies compared to local ones, since distant galaxies are more likely to be misclassified. This is not the case as can be seen on Fig. 4: the strong lensing data based on distant galaxies produce a similar $d(M/L)/d\varepsilon$ (although slightly smaller in average) than the data using the virial theorem based on local galaxies. More generally, an S0 contamination would induce a dependence of $d(M/L)/d\varepsilon$ with distance modulus. This was checked in Section 8.7. There is no evidence of increase of $d(M/L)/d\varepsilon$ with distance modulus.

**Correlation with apparent magnitude**

Similarly to the test previously discussed, an S0 contamination would make $d(M/L)/d\varepsilon$ to vary inversely to the apparent magnitude, since fainter galaxies are more difficult to properly classify. Since the apparent magnitude is tightly correlated with the distance modulus, the conclusion of the previous section should apply directly. However, we carried this check explicitly. Binning the $M/L$ from Bacon et al. (1985) in seven equidistant $B_t$ bins ($B_t$ is the integrated apparent blue magnitude), we extract $d(M/L)/d\varepsilon$ for each bin. A linear fit then yields $d(M/L)/d\varepsilon = (-0.97 \pm 1.97)B_t + 18.32 \pm 23.12$. There is no increase of $d(M/L)/d\varepsilon$ with $B_t$.

**Methodological check**

The agreement between $d(M/L)/d\varepsilon$ extracted with data using different technics for obtaining $M/L$ indicates that S0 contamination is not important: Technics involving detailed galaxy modelling are presumably more free of important misclassification but yield a similar $d(M/L)/d\varepsilon$, see Fig. 4.

While all these tests appear to directly or indirectly rule out any important bias from S0 contamination, they do not preclude a smaller contamination contribution to the observed correlation. Such possibility can be addressed only by individual dynamical studies of each galaxy.

## 9 CONCLUSION

Our analysis reveals a large correlation, with a slope: $\frac{d(M/L)}{d\varepsilon} = (14.53 \pm 4.77)\,M_\odot/L_\odot$, to be compared to our normalization of $M/L(r = R_{eff}, \varepsilon_{app} = 0.3) = 7.7\,M_\odot/L_\odot$. (The correlation slope scales with the normalization.) We have made a thorough investigation of the various interdependences of the many variables characterizing elliptical galaxies and found that they could not explain the observed correlation. (Of course, the possibility of a bias due to effects presently underappreciated cannot be definitely excluded.) In addition, the six methods that have been used to extract $M/L$ are independent. Consequently, to the best of our knowledge the correlation seems to be physical rather than a methodological, observational or instrumental bias. Its large magnitude makes it a feature that future descriptions of elliptical galaxies, which include their dark matter content, must be able to reproduce.

Numerically, with our normalization $M/L(r = R_{eff}, \varepsilon_{app} = 0.3) = 7.7\,M_\odot/L_\odot$, the correlation indicates that the most flattened medium-size elliptical galaxies have $M/L$ averaging to $(13.5 \pm 1.9)\,M_\odot/L_\odot$ and the round ones to $(3.3 \pm 1.5)\,M_\odot/L_\odot$. Since in the $B$ band, the stellar mass over luminosity is $M_*/L \simeq 4\,M_\odot/L_\odot$, this implies that within the radius investigated (typically 0.1–1 $R_{eff}$), round medium-size elliptical galaxies have small amounts of dark matter (including in particular those considered in Romanowsky et al. 2003). This is puzzling in light of the conventional model of cosmological structure formation, which requires dark matter seeds to trigger galaxy formations.

Finally, this attempt to understand an elusive quantity – dark matter in galaxies – may allow for the determination of another one perhaps even more evasive: the galaxy true ellipticity. As a practical application of the correlation, once the total mass has been assessed, the true ellipticity can be directly inferred.

## ACKNOWLEDGEMENTS

We acknowledge the help of G. Cates, K. Gebhardt, J. Gomez, L. V. E. Koopmans, A. M. Sandorfi and B. Terzić. This research used the NASA/IPAC Extragalactic Database (NED) operated by the Jet Propulsion Laboratory, California Institute of Technology, under contract with the National Aeronautics and Space Administration.

## REFERENCES

Arp H., 1966, Atlas of Peculiar Galaxies. Caltech, Pasadena, CA
Auger M. W., Treu T., Bolton A. S., Gavazzi R., Koopmans L. V. E., Marshall P. J., Bundy K., Moustakas L. A., 2009, ApJ, 705, 1099
Auger M. W., Treu T., Bolton A. S., Gavazzi R., Koopmans L. V. E., Marshall P. J., Moustakas L. A., Burles S., 2010, ApJ, 724, 511
Bacon R., Monnet G., Simien F., 1985, A&A, 152, 315
Bamford S. P. et al., 2009, MNRAS, 393, 1324
Barnabè M., Czoske O., Koopmans L. V. E., Treu T., Bolton A. S., Gavazzi R., 2009, MNRAS, 399, 21
Barnabè M., Czoske O., Koopmans L. V. E., Treu T., Bolton A. S., 2011, MNRAS, 415, 2215
Bender R., Surma P., Dobereiner S., Mollenhoff C., Madejsky R., 1989, A&A, 217, 35
Bertin G. et al., 1994, A&A, 292, 381
Bertola F., Bettoni D., Danziger J., Sadler E. M., Sparke L. S., de Zeeuw P. T., 1991, ApJ, 373, 369
Bertola F., Pizzella A., Persic M., Salucci P., 1993, ApJ, 416, L45
Binggeli B., 1994, in Meylan G., Prugniel Ph., eds, Proc. ESO/OHP Workshop on Dwarf Galaxies. ESO, Garching, p.13
Bolton A. S., Burles S., Koopmans L. V. E., Treu T., Gavazzi R., Moustakas L. A., Wayth R., Schlegel D. J., 2008, ApJ, 682, 964
Brighenti F., Mathews W., 1997, ApJ, 486 L83
Bruzual G., Charlot S., 2003, MNRAS, 344, 1000
Calvi R., Poggianti B. M, Fasano G., Vulcani B., 2012, MNRAS, 419, L14
Capaccioli M., Napolitano N. R., Arnaboldi M., 2003, in Semikhatov A., Vasilief M., Zalkin v., eds, Proc. The Third International Sakharov Conference on Physics, preprint (astro-ph/0211323)
Cappellari M. et al., 2006, MNRAS, 366, 1126
Cappellari M. et al., 2009, ApJ, 704, L34
Cappellari M. et al., 2011, MNRAS, 416, 1680
Cappellari M. et al., 2013a, MNRAS, 432, 1709
Cappellari M. et al., 2013b, MNRAS, 432, 1862
Cardone V. F., Tortora C., Molinaro R., Salzano V., 2009, A&A, 504, 769
Cardone V. F., Del Popolo A., Tortora C., Napolitano N. R., 2011, MNRAS, 416, 1822
Carollo C. M., Danziger I. J., 1994, MNRAS, 270, 523 (1994, MNRAS, 270, 743)
Carollo C. M., de Zeeuw P. T., van der Marel R. P., Danziger I. J., Qian E. E., 1995, ApJ, 441, 25
Chabrier G., 2003, PASP, 115, 763
Coccato L. et al., 2009, MNRAS, 394, 1249
Conroy C., van Dokkum P. G., 2012, ApJ, 760, 71
D'Onofrio M., Zaggia S. R., Longo G., Caon N., Capaccioli M., 1995, A&A, 296, 319

Das P., Gerhard O., Churazov E., Zhuravleva I., 2010, MNRAS, 409, 1362
Deason A. J., Belokurov V., Evans N. W., McCarthy I. G., 2012, ApJ, 748, 2
di Serego Alighieri S. et al., 2005, A&A, 442, 125
Djorgovski S., Davis M., 1987, ApJ, 313, 59
Dressler A., Lynden-Bell D., Burstein D., Davies R. L., Faber S. M., Terlevich R., Wegner G., 1987, ApJ, 313, 43
Driver S. P., Allen P. D., Liske J., Graham A. W., 2008, ApJ, 657, 85
Emsellem E. et al., 2011, MNRAS, 414, 888
Faber S. M., Jackson R. E., 1976, ApJ, 204, 668
Fassnacht C. D., Cohen J. G., 1998, ApJ, 115, 377
Faure C. et al., 2011, A&A, 529, A72
Ferreras I., Saha P., Williams L. R., 2005, ApJ 623, 5
Ferreras I., Saha P., Burles S., 2008, MNRAS, 383, 857
Foster C., Spitler L. R., Romanowsky A. J., Forbes D. A., Bekki K., Strader J., Proctor R. N., Brodie J. P., 2011, MNRAS, 415, 3393
Fukazawa Y., Botoya-Nonesa J. G., Pu J., Ohto A., Kawano N., 2006, ApJ, 636, 698
Gebhardt K. et al., 2003, ApJ, 583, 92
Graham A. W., Colless M. M., Busarello G., Zaggia S., Longo G., 1998, A&AS, 133, 325
Grillo C., Gobat R., Rosati P., Lombardi M., 2008, A&A, 477, L25
Grillo C., Gobat R., Lombardi M., Rosati P., 2009, A&A, 501, 461
Holden B. P. et al., 2005, ApJ, 620, 83
Humphrey P. J., Buote D. A., Gastaldello F., Zappacosta L., Bullock J. S., Brighenti F., Mathews W. G., 2005, in Wilson A., ed., Proc. 'The X-ray universe 2005' Symp., Dark Matter Halos of Early-Type Galaxies, available at: http://xmm.esac.esa.int/external/xmm_science/x-ray-symposium/
Humphrey P. J., Buote D. A., Gastaldello F., Zappacosta L., Bullock J. S., Brighenti F., Mathews W. G., 2006, ApJ, 646, 899
Jackson N., Helbig P., Browne I., Fassnacht C. D., Koopmans L., Marlow D., Wilkinson P. N., 1998, A&A, 334, L33
Jiang G., Kochanek C. S., 2007, ApJ, 671, 1568
Keeton C. R., Kochanek C. S., Falco E. E., 1998, ApJ, 509, 561
Kelson D. D., Illingworth G. D., van Dokkum P. G., Franx M., 2000, ApJ, 531, 184
Komatsu E. et al., 2011, ApJS, 192, 18
Koopmans L. V. E., Treu T., Bolton A. S., Burles S., Moustakas L. A., 2006, ApJ, 649, 599
Kormendy J., ApJ, 1977, 218, 333
Kronawitter A., Saglia R. P., Gerhard O., Bender R., 2000, A&A, 144, 53
Kroupa P., Tout C. A., Gilmore G., 1993, MNRAS, 262, 545
Lauer T. R., 1985, ApJ, 292, 104
Leier D., 2009, MNRAS, 400, 875
Leier D., Ferreras I., Saha P., Falco E. E., 2011, ApJ, 740, 97
Magorrian J., Ballantyne D., 2001, MNRAS, 322, 702
Magorrian J. et al., 1998, ApJ, 115, 2285
Maraston C., 2005, MNRAS, 362, 799
More A., Jahnke K., More S., Gallazzi A., Bell E. F., Barden M., Häußler B., 2011, ApJ, 734, 69
Nagino R., Matsushita K., 2009, A&A, 501, 157
Napolitano N. R. et al., 2005, MNRAS, 357, 691
Napolitano N. R. et al., 2009, MNRAS, 393, 329
NASA/IPAC Extragalactic Database, available at:http://nedwww.ipac.caltech.edu/
O'sulivan E., Sanderson A. J. R., Ponman T. J., 2007, MNRAS, 380, 1409
Pizzella A. et al., 1997, A&A, 323, 349
Proctor R. N., Forbes D. A., Romanowsky A. J., Brodie J. P., Strader J., Spolaor M., Mendel J. T., Spitler L., 2009, MNRAS, 398, 91
Prugniel Ph., Simien F., 1996, A&A, 309, 749
Rettura A. et al., 2006, A&A, 458, 717
Romanowsky A. J., Douglas N. G., Arnaboldi M., Kuijken K., Merrifield M. R., Napolitano N. R., Capaccioli M., Freeman K. C., 2003, Science, 301, 1696
Rowlands K. et al., 2012, MNRAS, 419, 2545
Ruff A. J., Gavazzi R., Marshall P. J., Treu T., Auger M. W., Brault F., 2011, ApJ, 727, 96
Saglia R. P., Bertin G., Stiavelli M., 1992, ApJ, 384, 433
Saglia R. P. et al., 1993, ApJ, 403, 567
Salpeter E. E., 1955, ApJ, 121, 161
Schwarzschild M., 1979, ApJ, 232, 236
Spiegel M. R., 1992, Theory and Problems of Probability and Statistics, 2nd edn. McGraw-Hill, New York
Statler T. S., 2001, ApJ, 121, 244
Thomas J., Saglia R. P., Bender R., Thomas D., Gebhardt K., Magorrian J., Corsini E. M., Wegner G, 2007, MNRAS, 382, 657
Thomas J. et al., 2011, MNRAS, 415, 545
Treu T., Koopmans L. V. E., 2004, ApJ, 611, 739
Treu T., Gavazzi R., Gorecki A., Marshall P. J., Koopmans L. V. E., Bolton A. S., Moustakas L. A., Burles S., 2009, ApJ, 690, 670
Trinchieri G., Fabbiano G., Kim D.-W., 1997, A&A, 318, 361
Tully R. B., Fisher J. R., 1977, A&A, 54, 661
van der Marel R. P., 1991, MNRAS, 253, 710
van der Marel R. P., van Dokkum P. G., 2007a, ApJ, 668, 738
van der Marel R. P., van Dokkum P. G., 2007b, ApJ, 668, 756
van der Wel A., Franx M., van Dokkum P. G., Rix H.-W., Illingworth G. D., Rosati P., 2005, ApJ, 631, 145
van Dokkum P. G., Stanford S. A., 2003, ApJ, 585, 78
Wegner G. A., Corsini E. M., Thomas J., Saglia R. P., Bender R., Pu S. D., 2012, ApJ, 144, 78

## APPENDIX A: TABLE OF RESULTS

**Notes for Table A1:**

(a) Added selection requirement that the galaxy characteristics listed in the reference are compatible with recent characteristics given in NED.

(b) Results using ellipticity correction from $\sigma^2$ isotropic method.

(c) Results using ellipticity correction from $\mu^2$ isotropic method.

(d) Results using ellipticity correction from $\mu^2$ anisotropic method.

(e) Distant galaxies (cluster CL1358+62). Added $M_B \leq -20$ selection to minimize the contamination from boxy galaxies.

(f) $M/L_B$ extracted for galactic cores.

(g) Distant galaxies.

(h) Also analysed with strong lensing method.

(i) $M_*$ obtained with composite stellar population models using a Kroupa IMF (Kroupa, Tout & Gilmore 1993).

(j) Two estimates of $M/L$ provided. One based on a two-integral Jeans model and one on a three-integral Schwartzchild model (Schwarzschild 1979).

(k) Kept LINERS, AGN and Seyfert galaxies since isothermal or virial equilibrium are not required.

(l) $\frac{M}{L}|_{sc}$ from Thomas et al. (2007, 2011) and Wegner et al. (2012) was assumed to be independent of radius.

(m) Homogenized compilation of literature (local galaxies).

(n) $\varepsilon_{\rm true}$ are from Statler (2001), Cappellari et al. (2006) and Foster et al. (2011).

(o) $a, b$ and $c$ are the radii of the galactic triaxial shape model.

(p) $q_0$ and $p_0$ are the intrinsic axis ratios of the triaxial model.

(q) Galaxies were identified so the standard $\sigma \geq 225\,{\rm km\,s^{-1}}$ selection for was not needed and not applied.

(r) Dark matter ratios extracted using either a Chabrier (Chabrier 2003) or a Salpeter IMF (Salpeter 1955).

(s) Use secondary infall model and Salpeter IMF.

(t) Stellar composite model used to obtain $M_*$ with two different sets of metallicity template Bruzual & Charlot (2003) or Maraston (2005) and three different IMF Salpeter (1955), Kroupa et al. (1993) or Chabrier (2003).

(u) Use results with adiabatic compression (favoured by the authors analysis).

**Table A1.** Column (1): publication reference. Column (2): method used to compute the dark matter content. Column 3): best-fitting result (not yet normalized to $M/L(\varepsilon_{\rm app}=0.3)=8\,{\rm M_\odot/L_\odot}$). Column (4): number of elliptical galaxies that passed the selection criteria given in Section 3. Column (5): reliability group, see Section 7.2, Column (6): weight factor, see Section 7.2. Column (7): specific notes regarding the analyses. The symbol '-' means that the data was not used in the global average.

| Ref. | Method | Best fit | Stat | rg | wf | Notes |
|---|---|---|---|---|---|---|
| Bacon et al. (1985) | Virial | $\frac{M}{L_B}=(13.08\pm2.97)\varepsilon_{\rm app}+(3.80\pm0.76)$ | 64 | 2 | 1.50 | a,b |
| Bacon et al. (1985) | Virial | $\frac{M}{L_B}=(5.91\pm4.67)\varepsilon_{\rm app}+(4.60\pm1.37)$ | 11 | 2 | 1.13 | a,c |
| Bacon et al. (1985) | Virial | $\frac{M}{L_B}=(6.19\pm3.59)\varepsilon_{\rm app}+(2.99\pm1.09)$ | 11 | 1 | 1.13 | a,d |
| Bender et al. (1989) | Virial | $\frac{M}{L_B}=(4.03\pm1.44)\varepsilon_{\rm app}+(2.17\pm0.43)$ | 35 | 3 | 2.37 | |
| Kelson et al. (2000) | Virial | $\frac{M}{L_V}=(15.53\pm7.21)\varepsilon_{\rm app}+(4.97\pm1.50)$ | 5 | 4 | 1.00 | e |
| Lauer (1985) | Virial | $\frac{M}{L_B}=(10.94\pm10.18)\varepsilon_{\rm app}+(12.89\pm3.01)$ | 10 | 4 | 6.09 | a,f |
| Leier (2009) | Virial | $\frac{M_{\rm vir}}{L_I(R_{\rm lense})}=(6.41\pm4.97)\varepsilon_{\rm app}+(-0.10\pm1.08)$ | 8 | 2 | 1.00 | g,h |
| Prugniel & Simien (1996) | Virial | $\frac{M}{L_B}=(6.58\pm1.98)\varepsilon_{\rm app}+(2.41\pm0.41)$ | 102 | 2 | 1.32 | |
| Rettura et al. (2006) | Virial | $\frac{M_{\rm dyn}}{M_*}=(4.29\pm1.36)\varepsilon_{\rm app}+(0.75\pm0.22)$ | 16 | 3 | 1.35 | g,i |
| van der Wel et al. (2005) | Virial | $\frac{M}{L_B}=(5.36\pm1.24)\varepsilon_{\rm app}+(1.07\pm0.24)$ | 13 | 3 | 1.35 | g |
| Cappellari et al. (2006) | Modelling | $\frac{M}{L}_{\rm Jeans}=(1.47\pm1.56)\varepsilon_{\rm true}+(2.37\pm0.55)$ | 6 | 1 | 1.38 | k |
| | | $\frac{M}{L}_{\rm Schw}=(1.09\pm1.68)\varepsilon_{\rm true}+(2.39\pm0.60)$ | 6 | 1 | 1.38 | |
| Cappellari et al. (2013a) | Modelling | $\frac{M}{L}=(2.02\pm1.48)\varepsilon_e+(3.51\pm0.42)$ | 32 | 1 | 1.86 | |
| Cappellari et al. (2013b) | Modelling | $\frac{M}{L}=(4.72\pm2.18)\varepsilon_e+(3.33\pm0.59)$ | 32 | 1 | 1.86 | |
| Kronawitter (2000) | Modelling | $\frac{M}{L_B}\|_{\rm in}=(4.58\pm7.04)\varepsilon_{\rm app}+(5.00\pm1.04)$ | 10 | 1 | 1.17 | l. |
| | | $\frac{M}{L_B}\|_{\rm out}=(-6.53\pm14.48)\varepsilon_{\rm app}+(7.79\pm2.22)$ | 10 | 1 | 1.17 | |
| Magorrian et al. (1998) | Modelling | $\frac{M}{L}=(-0.69\pm4.95)\varepsilon_{\rm app}+(5.43\pm1.90)$ | 7 | 1 | 1.33 | |
| Thomas et al. (2007, 2011) | Modelling | $\frac{M}{L}=(2.57\pm6.80)\varepsilon_{\rm true}+(6.25\pm1.75)$ | 7 | 1 | 1.16 | g,m |
| and Wegner et al. (2012) | | $\frac{M}{L}\|_{sc}=(-0.57\pm7.02)\varepsilon_{\rm true}+(7.24\pm1.71)$ | 7 | 2 | 1.63 | |
| van der Marel (1991) | Modelling | $\frac{M}{L_R}=(3.43\pm0.92)\varepsilon_{\rm app}+(2.22\pm0.11)$ | 9 | 1 | 1.47 | |
| van der Marel & van Dokkum (2007a) | Modelling | $\frac{M}{L_B}=(6.19\pm11.45)\varepsilon_{''true''}+(2.56\pm1.83)$ | 19 | 2 | 1.00 | g |
| van der Marel & van Dokkum (2007b) | Modelling | $\frac{M}{L_B}=(3.50\pm2.62)\varepsilon_{\rm app}+(6.09\pm0.75)$ | 17 | 1 | 2.04 | n |
| Capaccioli et al. (1992) | PNe/GC | $\frac{M}{L_B}=(19.43\pm7.85)\varepsilon_{\rm app}+(2.71\pm1.59)$ | 5 | | 2.14 | l |
| Deason et al. (2012) | PNe/GC | $\frac{M}{L}=(10.89\pm17.71)\varepsilon_{\rm app}+(10.56\pm5.87)$ | 7 | 1 | 1.24 | l |
| Magorrian & Ballantyne (2001) | PNe/GC | $\frac{M}{L}=(2.45\pm8.60)\varepsilon_{\rm app}+(6.73\pm1.20)$ | 6 | 1 | 1.14 | |
| Romanowsky et al. (2003) | PNe/GC | $\frac{M}{L_B}=(142.5\pm63.5)\varepsilon_{\rm true}+(-30.97\pm21.02)$ | 3 | 1 | 1.22 | o |
| Fukazawa et al. (2006) | X-ray | $\frac{M}{L_B}=(0.55\pm15.46)\varepsilon_{\rm app}+(4.74\pm3.98)$ | 7 | 2 | 1.17 | |
| Nagino & Matsushita (2009) | X-ray | $\frac{M}{L_B}(0.5R_{\rm eff})=(12.24\pm8.41)\varepsilon_{\rm app}+(8.71\pm2.56)$ | 3 | 2 | 1.17 | |
| | | $\frac{M}{L_B}(3R_{\rm eff})=(20.85\pm33.35)\varepsilon_{\rm app}+(17.47\pm10.46)$ | 3 | 2 | 1.17 | |
| | | $\frac{M}{L_B}(6R_{\rm eff})=(232.0\pm94.0)\varepsilon_{\rm app}+(-49.50\pm34.83)$ | 2 | 2 | 1.17 | |
| Bertola et al. (1991, 1993) | Gas disc | $\frac{M}{L_B}=(4.21\pm3.55)(1-\frac{bc}{a^2})+(1.62\pm1.61)$ | 4 | 1 | 1.15 | i,p |
| Pizzella et al. (1997) | Gas disc | $\frac{M}{L_T}=(1.39\pm13.13)(1-q_o p_o)+(3.37\pm6.60)$ | 4 | 1 | 1.15 | i,q |
| Auger et al. (2010) | Lensing | $\frac{M_{\rm tot}}{M_*}\|_{\rm Chab}=(3.38\pm0.79)\varepsilon_{\rm mass}+(1.55\pm0.15)$ | 34 | 1 | – | g,q,r |
| | | $\frac{M_{\rm tot}}{M_*}\|_{\rm Sal}=(1.86\pm0.46)\varepsilon_{\rm mass}+(0.89\pm0.09)$ | 34 | 1 | – | |
| | | $\frac{M_{\rm tot}}{M_*}\|_{\rm Chab}=(1.47\pm0.70)\varepsilon_{\rm app}+(1.82\pm0.16)$ | 34 | 1 | 2.04 | |
| | | $\frac{M_{\rm tot}}{M_*}\|_{\rm Sal}=(0.79\pm0.40)\varepsilon_{\rm app}+(1.05\pm0.09)$ | 34 | 1 | – | |
| Barnabè et al. (2011) | Lensing | $\frac{M}{L}\|_{\rm Chab.}=(33.1\pm16.7)\varepsilon_{\rm true}+(4.06\pm3.30)$ | 10 | 1 | 2.05 | g,q,r |
| | | $\frac{M}{L}\|_{\rm Sal}=(15.9\pm9.8)\varepsilon_{\rm true}+(3.07\pm1.86)$ | 10 | 1 | – | |
| Cardone et al. (2009) | Lensing | $\frac{M}{L}(R_{\rm eff})=(5.39\pm3.40)\varepsilon_{\rm mass}+(3.93\pm0.63)$ | 13 | 1 | – | g,q |
| | | $\frac{M}{L}(R_{\rm eff})=(1.94\pm3.52)\varepsilon_{\rm app}+(4.44\pm0.77)$ | 13 | 1 | 5.12 | |
| | | $\frac{M}{L}(R_{\rm Ein})=(4.57\pm2.79)\varepsilon_{\rm mass}+(3.81\pm0.56)$ | 13 | 2 | – | g,q,s |
| | | $\frac{M}{L}(R_{\rm Ein})=(2.50\pm3.00)\varepsilon_{\rm app}+(4.10\pm0.67)$ | 13 | 2 | 4.92 | |
| Cardone et al. (2011) | Lensing | $\frac{M}{L}(R_{\rm eff})=(8.17\pm1.60)\varepsilon_{\rm mass}+(4.83\pm0.27)$ | 36 | 2 | – | |
| | | $\frac{M}{L}(R_{\rm eff})=(4.82\pm1.32)\varepsilon_{\rm app}+(5.18\pm0.28)$ | 36 | 2 | 1.00 | |

**Table A1** – *continued*

| Ref. | Method | Best fit | Stat | rg | wf | Notes |
|---|---|---|---|---|---|---|
| Faure et al. (2011) | Lensing | $\frac{M_{\rm tot}}{M_*} = (30.21 \pm 6.13)\varepsilon + (-3.09 \pm 6.94)$ | 7 | 1 | 1.00 | g |
| Ferreras et al. (2005) | Lensing | $\frac{M_{\rm tot}}{M_*}\|_{V,{\rm Chab}} = (1.83 \pm 5.78)\varepsilon_{\rm app} + (0.83 \pm 4.45)$ | 4 | 1 | 1.29 | g,r |
| | | $\frac{M_{\rm tot}}{M_*}\|_{V,{\rm Sal}} = (3.16 \pm 3.69)\varepsilon_{\rm app} + (0.25 \pm 3.02)$ | 4 | 1 | – | |
| Ferreras et al. (2008) | Lensing | $\frac{M_{\rm tot}}{M_*} = (-0.47 \pm 2.63)\varepsilon_{\rm app} + (1.08 \pm 0.51)$ | 4 | 1 | 1.80 | g |
| | | $\frac{M_{\rm tot}}{M_*} = (-5.55 \pm 3.22)\varepsilon_{\rm mass} + (2.55 \pm 1.97)$ | 4 | 1 | – | |
| Grillo et al. (2009) | Lensing | $\frac{M_{\rm tot}}{L_B}\|_{\rm Mar,Sal} = (1.65 \pm 0.97)\varepsilon_{\rm app} + (1.66 \pm 0.22)$ | 40 | 1 | – | g,q,t |
| | | $\frac{M_{\rm tot}}{L_B}\|_{BC,{\rm Sal}} = (1.52 \pm 1.18)\varepsilon_{\rm app} + (1.74 \pm 0.26)$ | 40 | 1 | – | |
| | | $\frac{M_{\rm tot}}{L_B}\|_{\rm Mar,Krou} = (1.12 \pm 1.10)\varepsilon_{\rm app} + (1.83 \pm 0.25)$ | 40 | 1 | – | |
| | | $\frac{M_{\rm tot}}{L_B}\|_{BC,{\rm Chab}} = (1.75 \pm 1.06)\varepsilon_{\rm app} + (1.67 \pm 0.23)$ | 40 | 1 | 2.00 | |
| Jackson et al. (1998) | Lensing | $\frac{M_{\rm tot}}{L_H} = (0.80 \pm 1.67)\varepsilon_{\rm app} + (0.81 \pm 0.47)$ | 4 | 3 | 1.37 | g |
| Jiang & Kochanek (2007) | Lensing | $\frac{M_{\rm tot}}{M_*} = (1.24 \pm 0.92)\varepsilon_{\rm app} + (1.49 \pm 0.22)$ | 12 | 2 | 6.24 | g,q,u |
| | | $\frac{M_{\rm tot}}{M_*} = (0.14 \pm 0.68)\varepsilon_{\rm mass} + (1.74 \pm 0.19)$ | 12 | 2 | – | |
| Keeton et al. (1998) | Lensing | $\frac{M}{L_B} = (36.22 \pm 15.14)\varepsilon_{\rm mass} + (-5.64 \pm 5.34)$ | 3 | 3 | – | g,v |
| | | $\frac{M}{L_B} = (0.00 \pm 10.27)\varepsilon_{\rm app} + (6.55 \pm 3.64)$ | 3 | 3 | $\infty$ | |
| Koopmans et al. (2006) | Lensing | $\frac{M_{\rm tot}}{M_*} = (1.75 \pm 0.77)\varepsilon_{\rm app} + (0.92 \pm 0.16)$ | 9 | 1 | 2.52 | g,q |
| | | $\frac{M_{\rm tot}}{M_*} = (2.25 \pm 0.57)\varepsilon_{\rm mass} + (0.95 \pm 0.11)$ | 9 | 1 | – | |
| Leier (2009) | Lensing | $\frac{M}{L_I} = (5.51 \pm 3.27)\varepsilon_{\rm app} + (0.21 \pm 0.69)$ | 8 | 1 | 4.98 | g,q |
| Leier et al. (2011) | Lensing | $\frac{M}{L_V} = (-37.2 \pm 36.6)\varepsilon_{\rm app} + (24.02 \pm 10.60)$ | 3 | 2 | 2.83 | g |
| Ruff et al. (2011) | Lensing | $\frac{M}{M_*} = (-0.49 \pm 2.90)\varepsilon_{\rm app} + (0.71 \pm 0.76)$ | 7 | 2 | 1.00 | g,w |
| | | $\frac{M}{M_*} = (-0.53 \pm 2.47)\varepsilon_{\rm mass} + (0.71 \pm 0.63)$ | 7 | 2 | – | |
| | | $\frac{M_{\rm tot}}{M_*}\|_{\rm Sal} = (-3.62 \pm 8.81)\varepsilon_{\rm app} + (4.18 \pm 3.03)$ | 7 | 2 | – | |
| | | $\frac{M_{\rm tot}}{M_*}\|_{\rm Sal} = (4.66 \pm 1.43)\varepsilon_{\rm mass} + (1.17 \pm 0.61)$ | 7 | 2 | – | |
| | | $\frac{M_{\rm tot}}{M_*}\|_{\rm Chab} = (6.17 \pm 14.65)\varepsilon_{\rm app} + (3.46 \pm 4.10)$ | 7 | 2 | 1.00 | |
| | | $\frac{M_{\rm tot}}{M_*}\|_{\rm Chab} = (9.89 \pm 4.85)\varepsilon_{\rm mass} + (1.80 \pm 1.78)$ | 7 | 2 | – | |
| Treu & Koopmans (2004) | Lensing | $\frac{M}{L_B} = (2.04 \pm 3.20)\varepsilon_{\rm app} + (4.58 \pm 0.78)$ | 3 | 1 | 1.23 | g |

(v) We relaxed S0 rejection criteria and used $\sigma < 200\,{\rm km\,s^{-1}}$. Because it uses lenses that are all already employed in other analyses of higher reliability groups, this results is weighted out of the global average.

(w) The $M_*$ determined with a Chabrier or a Salpeter IMF are for $M/L$ at $R_{\rm eff}$. $M/L$ at the Einstein radius $R_{\rm Ein}$ does not need IMF input.

## APPENDIX B: INDIVIDUAL DATA SET PLOTS

Correlation between galactic dark matter content and ellipticity for the data sets used in our analysis. The corresponding best fits are shown by the straight lines and their values are given in Table A1. The results versus $\varepsilon_{\rm mass}$ or alternate IMF, which are not used for computing the final $d(M/L)/d\varepsilon$, are not shown.

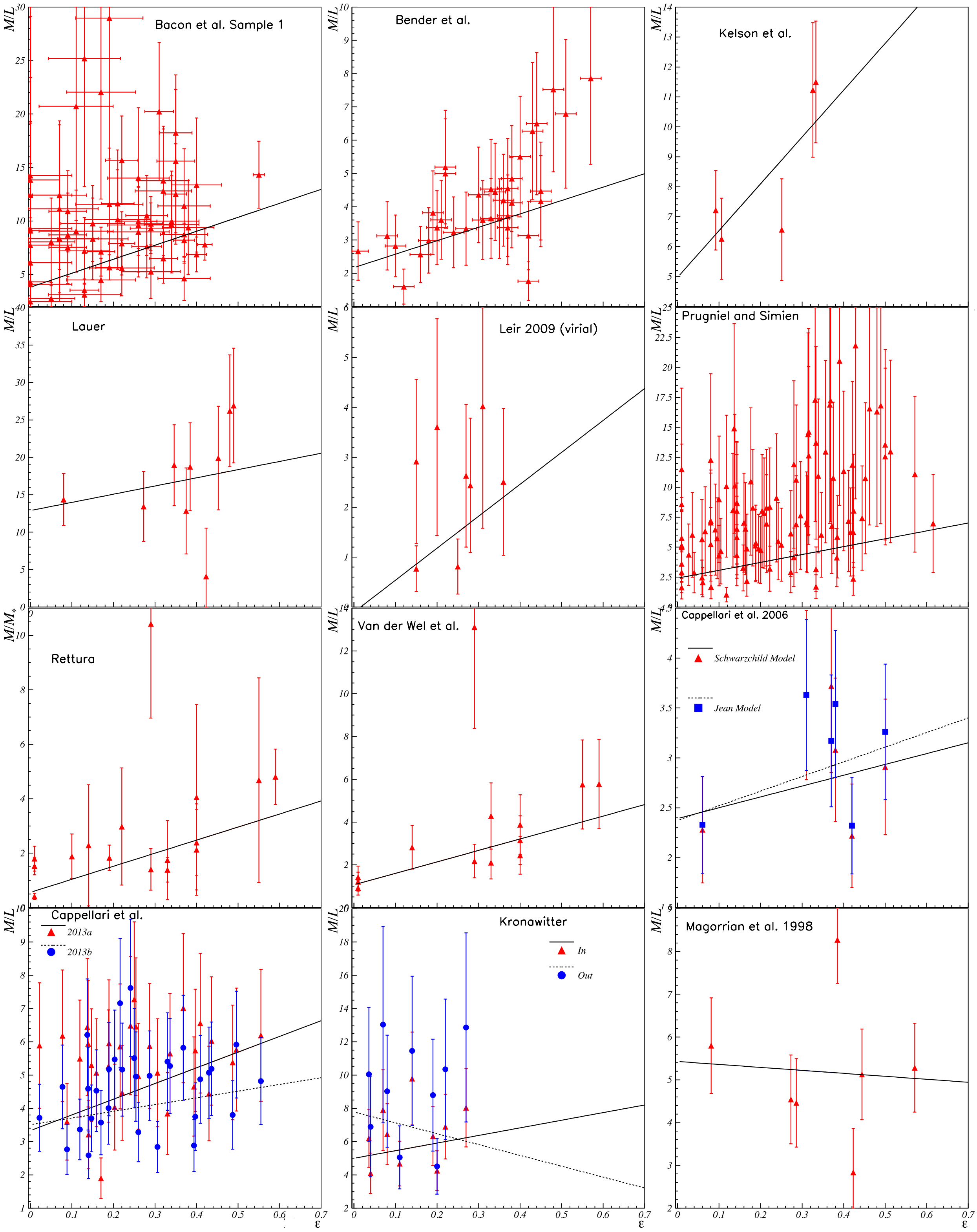

**Figure B1.** Galactic dark matter content versus ellipticity for, from top left to bottom right, Bacon et al. (1985), Bender et al. (1989), Kelson et al. (2000), Lauer (1985), Leier (2009), Prugniel & Simien (1996), Rettura et al. (2006), van der Wel et al. (2005), Cappellari et al. (2006, 2013a,b), Kronawitter (2000) and Magorrian et al. (1998).

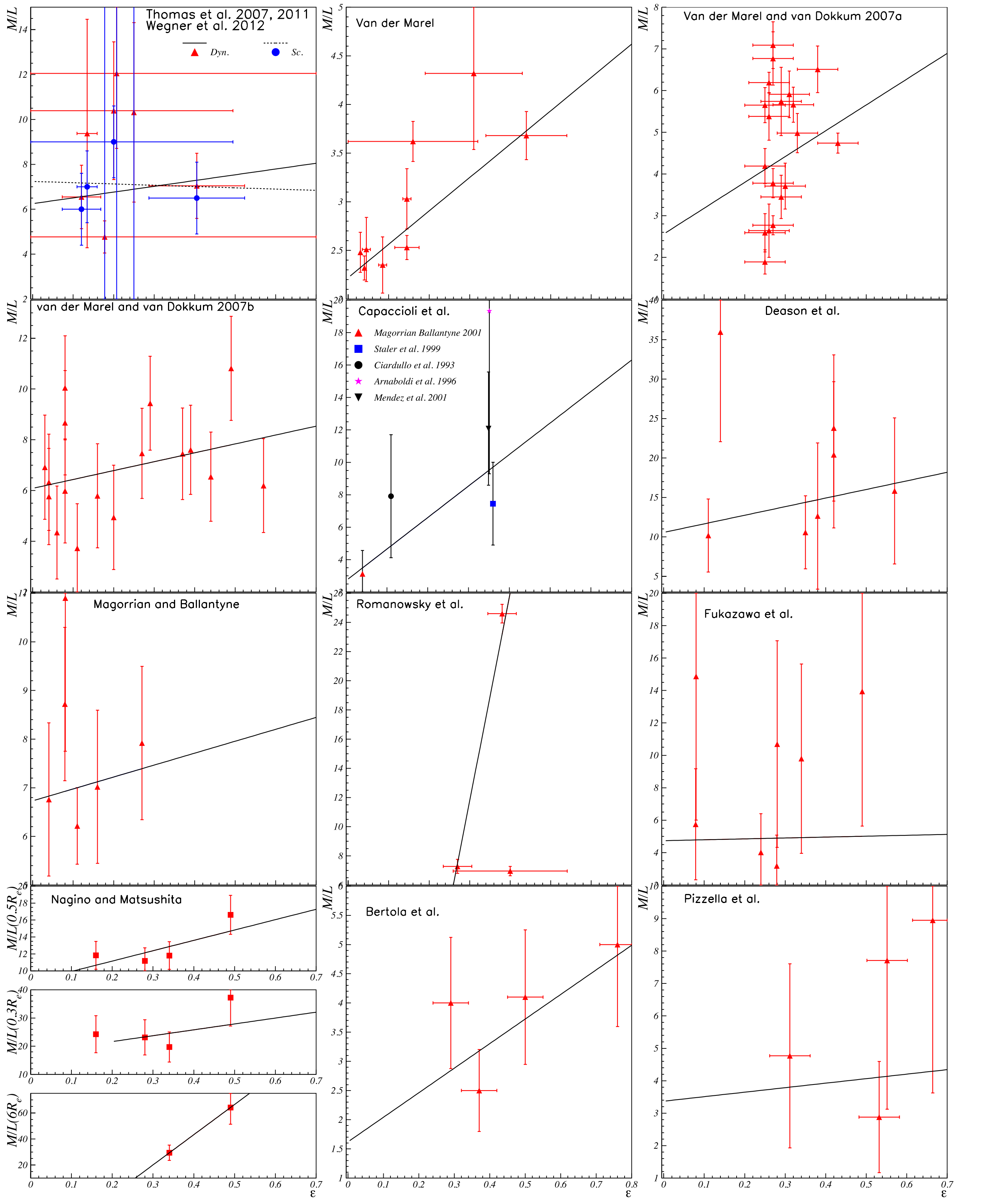

**Figure B2.** Galactic dark matter content versus ellipticity for, from top left to bottom right, Thomas et al. (2007, Thomas et al. 2011), Wegner et al. (2012), van der Marel (1991), van der Marel & van Dokkum (2007a, b), Capaccioli et al. (1992), Deason et al. (2012), Magorrian & Ballantyne (2001), Romanowsky et al. (2003), Fukazawa et al. (2006), Nagino & Matsushita (2009), Bertola et al. (1991, 1993) and Pizzella et al. (1997).

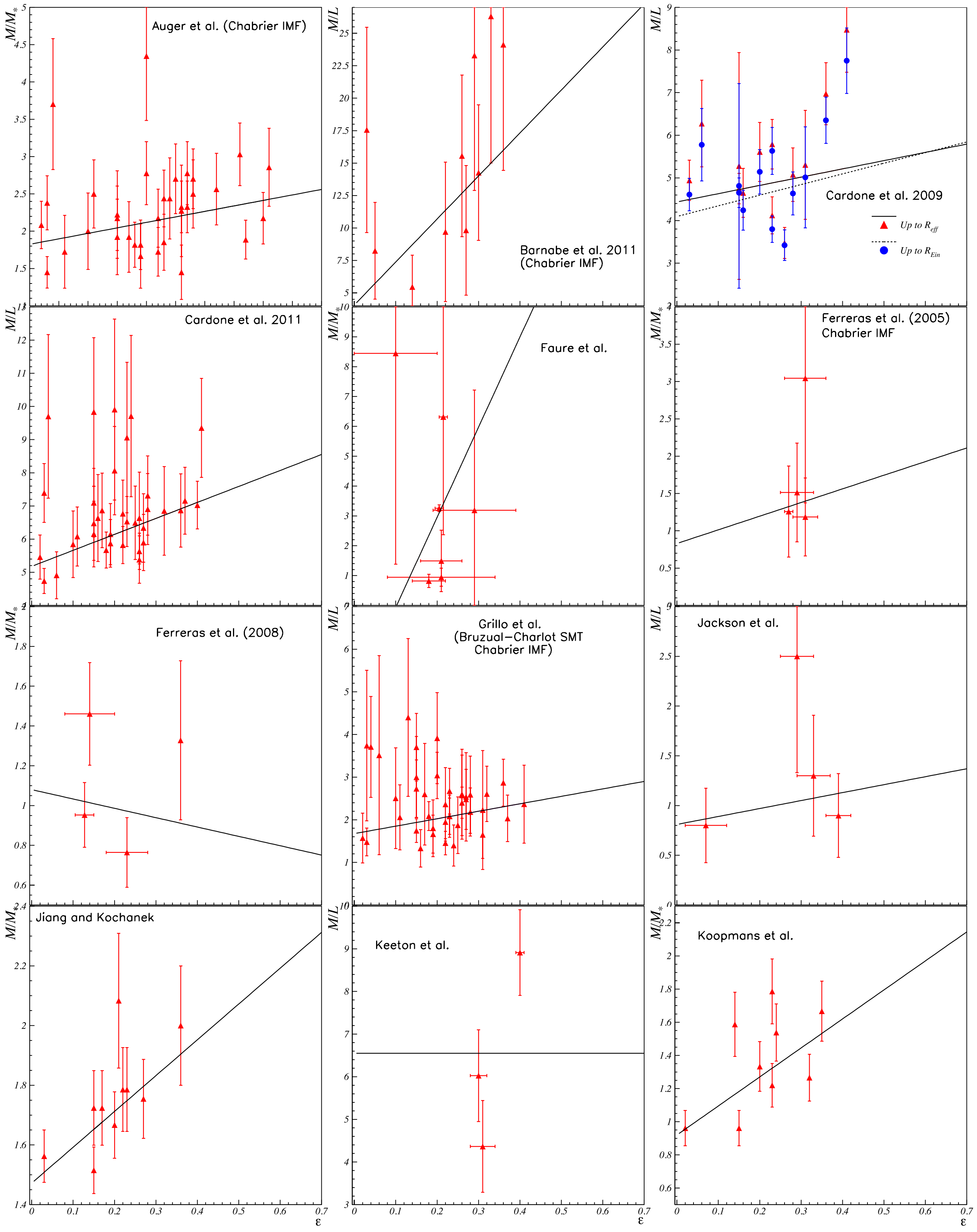

**Figure B3.** Galactic dark matter content versus ellipticity for, from top left to bottom right, Auger et al. (2010), Barnabè et al. (2011), Cardone et al. (2009, 2011), Faure et al. (2011), Ferreras et al. (2005, 2008), Grillo et al. (2009), Jackson et al. (1998), Jiang & Kochanek (2007), Keeton et al. (1998) and Koopmans et al. (2006).

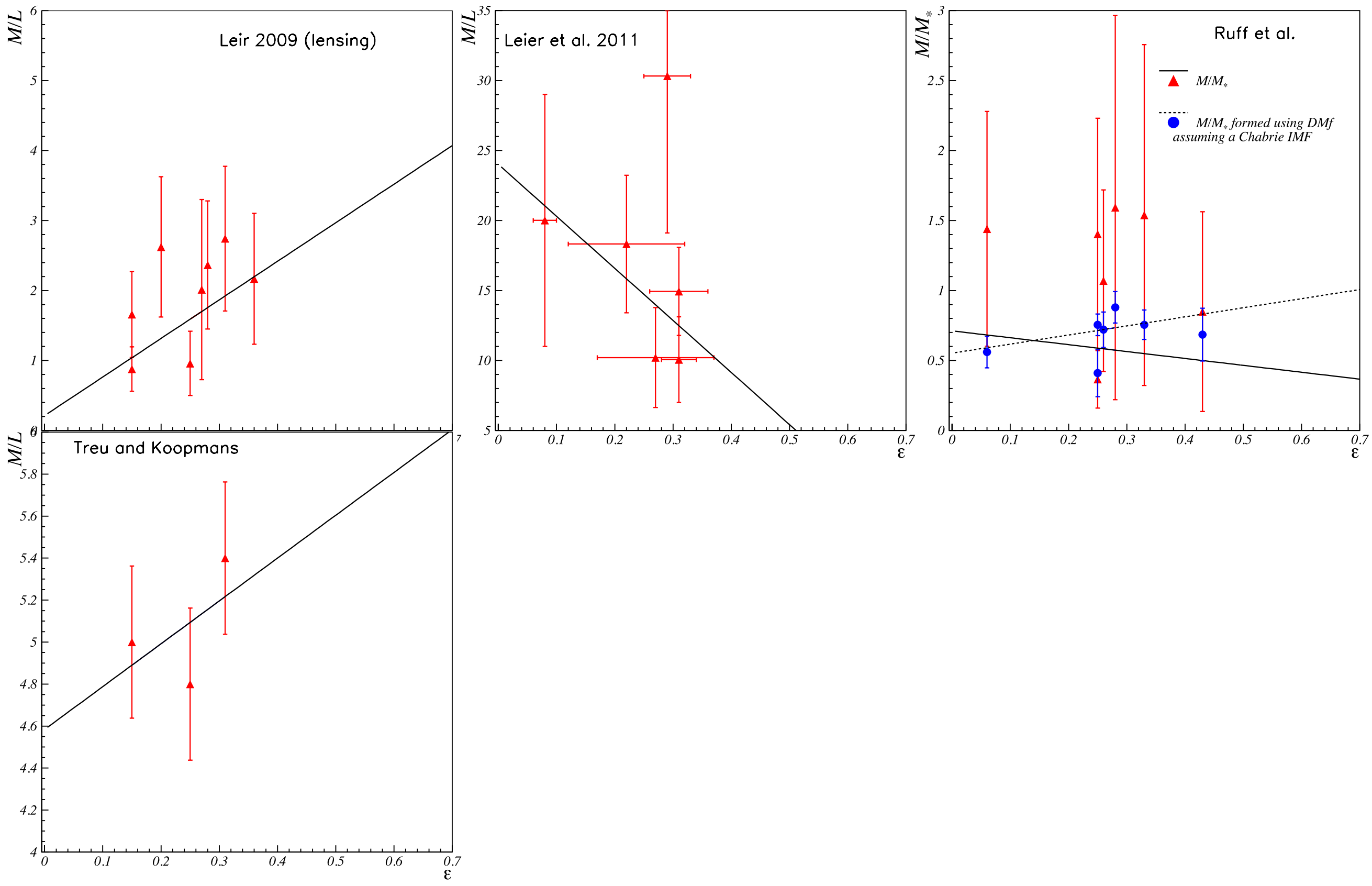

**Figure B4.** Galactic dark matter content versus ellipticity for, from top left to bottom right, Leier (2009), Leier et al. (2011), Ruff et al. (2011) and Treu & Koopmans (2004).

This paper has been typeset from a TEX/LATEX file prepared by the author.